\documentclass[a4paper,11pt]{article}
\pdfoutput=1 % if your are submitting a pdflatex (i.e. if you have
\usepackage{jheppub} % for details on the use of the package, please
\usepackage[T1]{fontenc} % if needed
\usepackage[left=1.9in,right=0.05in,top=2in,bottom = 0.5in]{geometry}
\usepackage{hyperref}
\usepackage{color}
\usepackage{amsmath}
\usepackage{amssymb} % lesssim
\usepackage{graphicx} % scalebox
\usepackage{mathrsfs} %\mathscr fonts
\usepackage[format=plain,labelformat=simple]{subcaption}

\usepackage{fancyhdr} % rhead; thispagestyle{fancy}
\usepackage{amsfonts}
\usepackage{tabularx} % multicolumn
\usepackage{multirow}
\usepackage{slashed}
\usepackage{xcolor}
\usepackage{booktabs} %thick table lines
\usepackage{bbm}
\usepackage{nicefrac} %command \nicefrac
\usepackage[
singlelinecheck=false % <-- important
]{caption}
\usepackage{feynmp} % Feynman Diagrams
\usepackage[utf8]{inputenc}
\def\be{\begin{equation}}
\def\ee{\end{equation}}
\def\bea{\begin{eqnarray}}
\def\eea{\end{eqnarray}}

\newcommand*{\rep}[2][]{\ensuremath{{\boldsymbol{#2}#1}}}

\newcommand{\U}[1]{\ensuremath{\mathrm{U}(#1)}}

\newcommand{\SU}[1]{\ensuremath{\mathrm{SU}(#1)}}

\definecolor{darkgreen}{HTML}{109930}

\title{\boldmath An $SU(5)\times U(1)^\prime$ SUSY GUT with a "vector-like chiral" fourth family to fit all low energy data, including the muon $g - 2$}

\author[a,1]{Harshal Kulkarni}
\author[b,2]{and Stuart Raby}

\affiliation[a]{Department of Physical Sciences, Indian Institute of Science Education and Research Kolkata, Mohanpur, West Bengal 741246, India}
\affiliation[b]{Department of Physics, The Ohio State University, Columbus, Ohio 43210, USA}

\emailAdd{hck18ms056@iiserkol.ac.in}
\emailAdd{raby.1@osu.edu}

\abstract{An additional generation of quarks and leptons and their SUSY counterparts, which are vector-like under the Standard Model gauge group but are chiral with respect to the new U(1)$_{3-4}$ gauge symmetry, are added to the Minimal Supersymmetric Standard Model (MSSM). We show that this model is a GUT and unifies the three SM gauge couplings and also the additional U(1)$_{3-4}$ coupling at a GUT scale of $\approx 5 \times 10^{16}$ GeV and explains the experimentally observed deviation of the muon $g-2$. We also fit the quark flavor changing processes consistent with the latest experimental data and look at the effect of the new particles on the $W$ boson mass without obviously conflicting with the observed masses of particles, CKM matrix elements, neutrino mixing angles, their mass differences, and the lepton-flavor violating bounds. This model predicts sparticle masses less than 25 TeV, with a gluino mass $\approx 2.3 - 3$ TeV consistent with constraints, and one of the neutralinos as the LSP with a mass of $\approx 480 - 580$ GeV, which is a potential dark matter candidate. The model is string theory motivated and predicts the VL quarks, leptons, a massive $Z'$ and two Dirac neutrinos at the TeV scale and the branching ratios of $\mu \longrightarrow e \gamma$, $\tau \longrightarrow \mu \gamma$ and $\tau \longrightarrow 3\mu$ with BR($\mu \longrightarrow e \gamma$) within reach of future experiments.}

\begin{document}
\maketitle
\flushbottom

\section{INTRODUCTION}
The Standard Model (SM) is very successful in predicting and explaining most of the experimental measurements. However, one of the discrepancies between the Standard Model and the experimental measurement that has been known for a long time is the muon anomalous magnetic moment. The discrepancy between the SM prediction and the measured value is \cite{Muong-2:2006rrc,Muong-2:2021ojo, Aoyama:2020ynm} :
\begin{equation}
    \Delta a_\mu = a_\mu^{exp} - a_\mu^{SM} = 251(59) \times 10^{-11}
\end{equation}
In addition to that, the $b \longrightarrow s l^+l^-$ transitions related to tests of lepton flavor non-universality in the observables $R(K)$ and $R(K^*)$ \cite{LHCb:2014cxe,LHCb:2014vgu}, semi-leptonic branching ratios \cite{LHCb:2013tgx,BaBar:2013qry,LHCb:2015wdu} and angular distributions \cite{LHCb:2013ghj,LHCb:2015svh,CMS:2017ivg,CMS:2015bcy,Belle:2016xuo,Belle:2016fev,ATLAS:2018gqc} also showed deviations from the SM predictions. However, recent LHCb measurements \cite{LHCb:2022zom} suggest that there is no lepton non-universality.

It has been shown that an extension of the SM involving VL leptons that couple only with the muons can explain the discrepancy of the muon $g$ - 2 \cite{Czarnecki:2001pv,Kannike:2011ng,Dermisek:2013gta}.
 It has also been shown that both the muon $g$ - 2 and the $b \longrightarrow s l^+l^-$ transitions related to tests of lepton flavor non-universality can be solved by introducing VL leptons, VL quarks and an additional massive $Z'$ boson corresponding to the spontaneously broken $U(1)_{\mu - \tau}$ \cite{Allanach:2015gkd,Altmannshofer:2016oaq,Megias:2017dzd}.  However, in light of the recent evidence for the absence of lepton flavor non-universality, it is most appropriate to fit all of the Wilson coefficients, resulting from beyond the Standard Model physics, $C_9$, $C_{10}$, $C_9'$ and $C_{10}'$ of the effective Hamiltonian \cite{Buras:1994dj,Bobeth:1999mk}:
\begin{equation}
    \mathcal{H}_{eff} = -\frac{4G_F}{\sqrt{2}}V_{tb}V_{ts^*} \frac{e^2}{16 \pi^2} \sum_{j = 9,10}(C_j \mathcal{O}_j + C'_j\mathcal{O}'_j) + h.c.
\end{equation}
where
\begin{equation}
    \mathcal{O}_9 = (\Bar{s}\gamma_\mu P_L b)(\Bar{\mu}\gamma^\mu \mu), \ \mathcal{O}'_9 = (\Bar{s}\gamma_\mu P_R b)(\Bar{\mu}\gamma^\mu \mu)
\end{equation}
and
\begin{equation}
    \mathcal{O}_{10} = (\Bar{s}\gamma_\mu P_L b)(\Bar{\mu}\gamma^\mu \gamma_5 \mu), \ \mathcal{O}'_{10} = (\Bar{s}\gamma_\mu P_R b)(\Bar{\mu}\gamma^\mu \gamma_5 \mu)
\end{equation}
to zero. This is what will be realized in our analysis, where we fit all the four Wilson coefficients within a value of $\pm$ 0.1.

In addition, it was shown in \cite{Raby:2017igl} that an additional 4th family which is VL with respect to the SM but chiral with respect to the new spontaneously broken U(1)$_{3-4}$ symmetry solves the muon $g$ - 2 anomaly without directly conflicting with experimental data. The analysis performed there, however, is restricted to the second and third  families and is non-supersymmetric. However, since string theory gives a supersymmetric theory,  in this paper, we generalize to include supersymmetry and present a complete analysis involving the three SM families and a VL-chiral family throughout. Finally, experimental evidence \cite{CDF:2022hxs, ATLAS:2023fsi} shows a deviation of the $W$ boson mass from its value predicted by the SM. It was shown in \cite{Kawamura:2022fhm} that the recent measurement \cite{CDF:2022hxs} of the $W$ boson mass can be fit by considering VL leptons and an additional U(1) gauge group. In our analysis, we extend the model of \cite{Raby:2017igl} and also study the effect of the new vector-like chiral leptons of our model on the $W$ boson mass.

In our model, the third SM family, the left-handed VL family and their SUSY counterparts have charges +1 and -1 respectively under the new U(1)$_{3-4}$ gauge group and all other fermions and their sparticles are neutral. This model is motivated by heterotic string orbifold constructions \cite{Buchmuller:2005jr,Buchmuller:2006ik,Lebedev:2006kn,Lebedev:2007hv,Lebedev:2008un,Blaszczyk:2009in,Kappl:2010yu} which give the full MSSM spectrum, states which are VL under the SM gauge group and chiral under new $U(1)'$ gauge symmetries. In addition, there are many SM singlet scalars that break the additional gauge symmetries and give mass to the new vector bosons and the VL particles. In many previous constructions, the additional particles gained mass at the string scale; however, in this model, we consider the case when at least one of the families of the VL particles is light enough ($\sim$ 1 TeV), hence making it possible to observe them in future experiments. We consider the Lagrangian involving this additional VL family and the corresponding sparticles, but the higher mass modes predicted by string theory are assumed to be integrated out.

It will be seen that the model simultaneously fits the observed quark and lepton masses of the three SM families, the Higgs mass, the strong and U(1)$_{em}$ gauge couplings, the CKM matrix elements, and neutrino mass differences and mixing angles. It also fits the muon anomalous magnetic moment and the deviation of the $W$ boson mass from its SM value \cite{ParticleDataGroup:2016lqr} without violating the data from quark flavor changing processes (i.e. Lepton Flavor Violating (LFV) decay bounds) and the cosmological upper bound on the sum of neutrino masses, i.e. $\sum_{i = 1}^3m_{\nu_i} < 0.12$ eV. The new VL leptons, VL quarks, Dirac neutrinos as well as the $Z'$ boson all have mass
 at the TeV scale. The model being supersymmetric predicts all the sparticle masses less than 25 TeV and also gives a $\approx 480 - 580$ GeV neutralino, which is the lightest supersymmetric particle (LSP) of the model and has the potential to be a dark matter candidate.

\section{MODEL}
The model under consideration is the MSSM with three right-handed neutrinos, one additional generation of left-handed and right-handed chiral fields, and their supersymmetric counterparts -- which correspond to the VL particles. Further, we also have an additional $U(1)_{3-4}$ under which the third SM family and the left-handed chiral part of the fourth additional generation are charged. This $U(1)_{3-4}$ gauge symmetry is spontaneously broken by the nonzero vacuum expectation value (VEV) of the scalar components of the superfields $\Phi$ and $\bar \Phi$. All the relevant scalar, fermionic fields of the model and their quantum numbers are given in terms of the superfields in the \textbf{Table 1} below.\footnote{The additional $U(1)$ gauge interaction commutes with $SU(5)$ (and even $SO(10)$ which is not gauged). It is flavor dependent and might result in string theory.}

\begin{table}[!htbp]
  \begin{center}
    \begin{tabular}{lccccccc}
      \toprule
      $G_\mathrm{SM}$ family & $\left(\rep{3},\rep{2}\right)_{\frac16}$ & $\left(\rep{3},\rep{1}\right)_{-\frac23}$ & $\left(\rep{3},\rep{1}\right)_{\frac13}$ & $\left(\rep{1},\rep{2}\right)_{-\frac12}$ & $\left(\rep{1},\rep{1}\right)_{1}$ & $\left(\rep{1},\rep{1}\right)_{0}$ & $\U1_{3-4}$ \\
     \midrule
      $a=1,2$ &$q^{a}=\left(u^a,d^a\right)$ & $\bar u^a$ & $\bar d^a$ & $l^a=\left(\nu^a,e^a\right)$ & $\bar e^a$ & $\bar \nu^a$ & $0$ \\
      $3$ & $q^3=\left(u^3,d^3\right)$ & $\bar U$  & $\bar D$  & $l^3=\left(\nu^3,e^3\right)$ & $\bar E$ & $\bar N$ & $1$ \\
        $4$ & \hspace{1pt}$Q =\left(U',D'\right)$ & $\bar u^3$ & $\bar d^3$ & \;\hspace{1pt}$L =\left(N',E'\right)$ & $\bar e^3$  & $\bar \nu^3$  & $-1\phantom{-}$ \\
        \toprule
        $G_\mathrm{SM}$ family & $\left(\rep{\bar 3},\rep{\bar 2}\right)_{-\frac16}$ & $\left(\rep{\bar 3},\rep{1}\right)_{\frac23}$ & $\left(\rep{\bar 3},\rep{1}\right)_{-\frac13}$ & $\left(\rep{1},\rep{\bar 2}\right)_{\frac12}$ & $\left(\rep{1},\rep{1}\right)_{-1}$ & $\left(\rep{1},\rep{1}\right)_{0}$ & $\U1_{3-4}$ \\  \midrule
      $\bar 4$ & \,$\bar Q =\left(\bar U',\bar D'\right)$ & $U$ & $D$ & \;\,$\bar L =\left(\bar N',\bar E'\right)$ & $E$ & $N$ & $0$\\
      \bottomrule
    \end{tabular}
    \\[0.3cm]
    \begin{tabular}{lcccc}
      \toprule
      & $H_u$ & $H_d$ & $\Phi$ & $\bar \Phi$  \\
      \midrule
      $G_\mathrm{SM}$ & $\left(\rep{1},\rep{2}\right)_{\frac12}$ & $\left(\rep{1},\rep{2}\right)_{-\frac12}$ & $\left(\rep{1},\rep{1}\right)_{0}$ & $\left(\rep{1},\rep{1}\right)_{0}$  \\
      $\U1_{3-4}$ & 0 & 0 & 1 & -1  \\
      \bottomrule
    \end{tabular}
  \caption{The quantum numbers of fermions and scalars in our model under the SM gauge group and under the new $\U1'\equiv\U1_{3-4}$. Note that
  the `third' and left-handed `fourth' family are actually mixtures of the actual third and left-handed fourth family.  This is necessary to
  allow for a dimension 4 Yukawa coupling for the third family of quarks and leptons. In fact, in terms of Dirac notation the third family right-handed states have the same  U(1)$_{3-4}$ charge as the left-handed quark and lepton doublets. The same is true for the 4th family right-handed states.
  Note, in addition, that primed fields have nothing to do with the $\U1'$ \textit{per se} but are used to denote constituents of $\SU2_\mathrm{L}$ doublets.
  }\label{tab:SM4}
  \label{tab:qm}
  \end{center}
\end{table}

The superpotential relevant for our analysis is:
\begin{align*}
    \mathcal{W} &= - \Bar{u}_{i}y^u_{ij}q_{j}H_u - \Bar{d}_{i}y^d_{ij}q_{j}H_d - \Bar{e}_{i}y^e_{ij}l_{j}H_d - \Bar{\nu}_{i}y^\nu_{ij}l_{j}H_u \\
    & - \lambda_{LRU} \Bar{U} Q H_u -  \lambda_{LRD} \Bar{D} Q H_d - \lambda_{LRE} \Bar{E} L H_d - \lambda_{LRN} \Bar{N} L H_u \\
    & - \lambda_{RLU} \Bar{Q} U H_d -  \lambda_{RLD} \Bar{Q} D H_u - \lambda_{RLE} \Bar{L} E H_u - \lambda_{RLN} \Bar{L} N H_d \\
    & - \bar \Phi ( \lambda_{3Q} q^3 \Bar{Q} + \lambda_{U} \Bar{U} U + \lambda_{D} \Bar{D} D + \lambda_{3L} l^3 \Bar{L} + \lambda_{E} \Bar{E} E + \lambda_{N} \Bar{N} N )\\
    &- \Phi (\lambda_Q Q \Bar{Q} + \lambda_{3U} \Bar{u}^3 U + \lambda_{3D} \Bar{d}^3 D + \lambda_L L \Bar{L} + \lambda_{3E}  \Bar{e}^3 E + \lambda_{3N} \Bar{\nu}^3 N) \\
     &- \lambda_{2Q_a}\Bar{Q} q^a - \lambda_{2L_a}\Bar{L} l^a - \lambda_{2U_a}\Bar{u}^a U - \lambda_{2D_a}\Bar{d}^a D
    - \lambda_{2E_a}\Bar{e}^a E - \lambda_{2N_a}\Bar{\nu}^a N - \mu H_u H_d - \mu_\Phi \bar \Phi \Phi
\end{align*}
where $i,j = 1,2,3$ and the Yukawa matrices $y^u_{ij}, y^d_{ij}, y^e_{ij}$ and $y^\nu_{ij}$ are block diagonal due to the constraint of the Lagrangian being invariant under the $U(1)_{3-4}$ gauge symmetry. Almost all of the Yukawa parameters are taken to be real. The Yukawa couplings $y^d_{12}$, $y^d_{21}$, $y^d_{22}$, $y^u_{33} = y^d_{33} = y^e_{33}$ (in the GUT sense) and $\lambda_{3Q}$, $\lambda_{3U}$ and $\lambda_{3D}$ are taken to be complex to get the complex parameters in the CKM matrix. The \textbf{2} $\oplus$ \textbf{1} flavor structure of the SM families can, in principle, originate from a $D_4$ flavor symmetry \cite{Kobayashi:2004ud,Kobayashi:2004ya,Blaszczyk:2009in,Kappl:2010yu,Kobayashi:2006wq,Ko:2007dz}, however this is beyond the scope of the present paper.  The quantities $\{\lambda_{2Q_a}, \lambda_{2L_a}, \lambda_{2U_a}, \lambda_{2D_a}, \lambda_{2E_a}, \lambda_{2N_a} \}$ are effective $\mu$-like parameters with mass dimension $1$. We will present a detailed analysis of the flavor physics of all the three SM families. \\

We consider the generic soft SUSY breaking Lagrangian:
\begin{equation}
    \mathcal{L}_{soft} = -\frac{1}{6}h^{ijk}\phi_i \phi_j \phi_k - \frac{1}{2}b^{ij} \phi_i \phi_j - \frac{1}{2} (m^2)^i_j \phi^{*j}\phi_i - \frac{1}{2}M_a \lambda_a \lambda_a
\end{equation}
where the scalar fields $\phi_i$ are the sparticle fields, the Higgs and new scalar fields of our model and the $\lambda_a$ are the four gaugino fields. The RG flow of the soft SUSY breaking sparticle and gaugino mass parameters have also been analyzed. We also consider Majorana neutrino mass terms in the superpotential, given by the addition of:
\begin{equation}
\mathcal{W}_{maj} = \frac{1}{2} M_L N N + \frac{1}{2} M_R^{ab} \bar \nu^a \bar \nu^b + M_R \bar N \bar \nu^3 .
\end{equation}
This then gives the Lagrangian terms for the fermions (in Dirac notation)
\begin{equation}
    \mathcal{L}_{maj} = -\frac{1}{2}M_L \overline{N_L^\mathcal{C}} N_L -\frac{1}{2} M_R^{ab}\overline{\left( \bar{\nu}_L^{a}\right)^\mathcal{C}} \bar{\nu}_L^b - \left(M_R \overline{\bar{N}_L^\mathcal{C}} \bar{\nu}_L^3+\mathrm{h.c.}\right) + \mathrm{h.c.}
\end{equation}
We take the Majorana mass terms ${M_L, M_R, M_R^{ab}}$ to be much larger than the VEVs  ${v_\Phi, v_{\bar{\Phi}} >  v_{H_u}, v_{H_d}}$. Hence, we can integrate out the heavy states. The details of this calculation are presented in \hyperref[sec:7.1]{Appendix ~\ref*{sec:7.1}}. After integrating out the heavy states, we get an effective Lagrangian given by:
 \begin{align*}
    \mathcal{L}^{eff} &=  \overline{(l^{d}_L)^C} (y^\nu_{d e})^T H_u (M_R^{-1})_{e a} y^\nu_{db} l^b_{L}H_u  +  \frac{1}{M_R} \lambda_{LRN}  (y^\nu_{33}) \overline{(l^{3}_L)^C} H_u  L_L H_u  \\ & - \overline{{\bar L}_L^C} \left(  \lambda_{2L_a} l_L^a + \lambda_{3 L} \bar \Phi l^3_L + \lambda_L \Phi L_L   \right) +   \frac{1}{M_L} \lambda_{RLN} \lambda_{RLN} \overline{{\bar L}_L^C} H_d  \bar{L}_L H_d  + \mathrm{h.c.}
 \end{align*}
This effective Lagrangian is considered for RG flow and for the physics below the right-handed Majorana neutrino mass scale.\\
At the weak scale,  the scalar fields obtain vacuum expectation values and the above effective Lagrangian becomes:
\begin{align*}
    \mathcal{L}^{eff} &=  \overline{(l^{d}_L)^C} ~(y^\nu_{d e})^T (M_R^{-1})_{e a} y^\nu_{db} ~l^b_{L} ~v_{H_u}^2  +  \frac{1}{M_R} \lambda_{LRN}  y^\nu_{33}  ~\overline{(l^{3}_L)^C}  ~L_L ~v_{H_u}^2   \\ & - \overline{({\bar N^{\prime}_L})^C} \left(  \lambda_{2L_a} \nu_L^a + \lambda_{3 L} v_{\bar \Phi} \nu^3_L + \lambda_L v_{\Phi} ~N^\prime_L \right) - \bar E^\prime_R \left(  \lambda_{2L_a} e_L^a + \lambda_{3 L} v_{\bar \Phi} e^3_L + \lambda_L v_{\Phi} E^\prime_L   \right) \\ &+ \frac{1}{M_L} \lambda_{RLN} \lambda_{RLN} \overline{{\bar L}_L^C} \bar{L}_L ~v_{H_d}^2 + \mathrm{h.c.}
\end{align*}
There are only 5 light neutrino states at the weak scale. Of these, one linear combination of $\{ \nu_L^1, ~\nu_L^2, ~\nu_L^3, ~N^\prime_L\}$ obtains mass at the heavier scale $\{ \sim v_\Phi, v_{\bar \Phi} \}$.  This is a Dirac mass term and includes two Weyl spinors. The other 3 linear combinations only obtain mass at the scale of order $v_{H_u}^2/M_R$, etc.

\subsection{Quark and charged lepton masses}
We can write the 5 $\times$ 5 Dirac mass matrices as:
\begin{equation}
    \Bar{e}_R \mathcal{M}^e e_L = (\Bar{e}_{Ri} ~\Bar{E}_R ~\Bar{E}'_R) \begin{pmatrix}
v_{H_d}y^e_{11} & v_{H_d}y^e_{12} & 0 & 0 & \lambda_{2E_1}\\
v_{H_d}y^e_{21} & v_{H_d}y^e_{22} & 0 & 0 & \lambda_{2E_2}\\
0 & 0 & v_{H_d}y^e_{33} & 0 & v_{\Phi}\lambda_{3E} \\
0 & 0 & 0 & v_{H_d}\lambda_{LRE} & v_{\bar \Phi}\lambda_{E} \\
\lambda_{2L_1} & \lambda_{2L_2} & v_{\bar \Phi}\lambda_{3L} & v_{\Phi}\lambda_L & v_{H_u}\lambda_{RLE}
\end{pmatrix}
\begin{pmatrix}
e_{Li} \\
E'_L \\
E_L
\end{pmatrix}
\end{equation}
\begin{equation}
    \Bar{u}_R \mathcal{M}^u u_L = (\Bar{u}_{Ri} \Bar{U}_R \Bar{U}'_R) \begin{pmatrix}
v_{H_u}y^u_{11} & v_{H_u}y^u_{12} & 0 & 0 & \lambda_{2U_1}\\
v_{H_u}y^u_{21} & v_{H_u}y^u_{22} & 0 & 0 & \lambda_{2U_2}\\
0 & 0 & v_{H_u}y^u_{33} & 0 & v_{\Phi}\lambda_{3U} \\
0 & 0 & 0 & v_{H_u}\lambda_{LRU} & v_{\bar \Phi}\lambda_{U} \\
\lambda_{2Q_1} & \lambda_{2Q_2} & v_{\bar \Phi}\lambda_{3Q} & v_{\Phi}\lambda_Q & v_{H_d}\lambda_{RLU}
\end{pmatrix}
\begin{pmatrix}
u_{Li} \\
U'_L \\
U_L
\end{pmatrix}
\end{equation}
\begin{equation}
    \Bar{d}_R \mathcal{M}^d d_L = (\Bar{d}_{Ri} \Bar{D}_R \Bar{D}'_R) \begin{pmatrix}
v_{H_d}y^d_{11} & v_{H_d}y^d_{12} & 0 & 0 & \lambda_{2D_1}\\
v_{H_d}y^d_{21} & v_{H_d}y^d_{22} & 0 & 0 & \lambda_{2D_2}\\
0 & 0 & v_{H_d}y^d_{33} & 0 & v_{\Phi}\lambda_{3D} \\
0 & 0 & 0 & v_{H_d}\lambda_{LRD} & v_{\bar \Phi}\lambda_{D} \\
\lambda_{2Q_1} & \lambda_{2Q_2} & v_{\bar \Phi}\lambda_{3Q} & v_{\Phi}\lambda_Q & v_{H_u}\lambda_{RLD}
\end{pmatrix}
\begin{pmatrix}
d_{Li} \\
D'_L \\
D_L
\end{pmatrix}
\end{equation}
where we have assumed the scalar VEVs to be real. The weak scale Yukawa matrices and the dimension 1 parameters being used in the above computation are given in \hyperref[sec:7.5]{Appendix~\ref*{sec:7.5}} and \hyperref[sec:7.8]{Appendix~\ref*{sec:7.8}} for the two fit points. The VEVs $v_{H_u}$ and $v_{H_d}$ are related to the Higgs VEV by:
\begin{equation}
    \frac{v^2}{2} = v_{H_u}^2 + v_{H_d}^2
\end{equation}
and:
\begin{equation}
    tan\beta = \frac{v_{H_u}}{v_{H_d}}
\end{equation}
These together imply:
\begin{equation}
    v_{H_d} = \frac{v}{\sqrt{2(1+tan^2\beta)}}, \ v_{H_u} = \frac{v~tan\beta}{\sqrt{2(1+tan^2\beta)}}
\end{equation}
Denoting the electrically charged fermions by $f = e, u, d$, the mass basis is defined as:
\begin{equation}
    \hat{f}_L = (U^f_L)^{\dag} f_L, \ \hat{f}_R = (U^f_R)^{\dag} f_R
\end{equation}
The unitary matrices $U^f_L$ and $U^f_R$ diagonalize the Dirac mass matrices:
\begin{equation}
    (U^e_R)^\dag \mathcal{M}^e U^e_L = diag(m_e, m_\mu, m_\tau, m_{E_1}, m_{E_2})
\end{equation}
\begin{equation}
    (U^u_R)^\dag \mathcal{M}^u U^u_L = diag(m_u, m_c, m_t, m_{U_1}, m_{U_2})
\end{equation}
\begin{equation}
    (U^d_R)^\dag \mathcal{M}^d U^d_L = diag(m_d, m_s, m_b, m_{D_1}, m_{D_2})
\end{equation}
where the $m_{E_a}$, $m_{U_a}$ and $m_{D_a}$ for $a = 1,2$ are the masses of the new charged fermions due to the introduction of a 4th vector-like chiral family. The masses are arranged in increasing order.

\subsection{The Extended CKM matrix}
From the $U^f_L$ and $U^f_R$ obtained after diagonalization, the extended CKM matrix can be given by:
\begin{equation}
    \Hat{V}_{CKM} = (U^u_L)^\dag P_{\bar{5}} U^d_L
\end{equation}
where $P_{\bar{5}} = diag(1,1,1,1,0)$ is a projection operator. As can be clearly seen from above, the extended CKM matrix is not unitary. The 3 $\times$ 3 submatrix of this extended CKM matrix is the SM CKM matrix.

We will show that the point in parameter space that we have considered will, within error, reproduce the observed quark and charged lepton masses for the 3 SM families and the CKM matrix. The weak scale CKM matrix for the two fit points can be found in \hyperref[sec:7.6]{Appendix~\ref*{sec:7.6}} and \hyperref[sec:7.9]{Appendix~\ref*{sec:7.9}}.

\subsection{Neutrino mixing angles and mass differences}
From the effective Lagrangian at the weak scale mentioned earlier, the mass matrix for the light neutrinos is given by:
\begin{equation}
\footnotesize{ \left( \begin{array}{ccccc} \overline{\nu_L^{1 C}} & \overline{\nu_L^{2 C}}  & \overline{\nu_L^{3 C}}  & \overline{N_L^{\prime C}}  & \overline{\bar N_L^{\prime C}} \end{array} \right) \left( \begin{array}{ccccc}
   ( y^{\nu T} (M_R^{ab})^{-1} y^\nu )_{11} v_u^2 & ( y^{\nu T} (M_R^{ab})^{-1} y^\nu )_{12} v_u^2 & 0 & 0 &  \lambda_{2L_1} \\
   ( y^{\nu T} (M_R^{ab})^{-1} y^\nu )_{21}  v_u^2& ( y^{\nu T} (M_R^{ab})^{-1} y^\nu )_{22} v_u^2 & 0 & 0 &  \lambda_{2L_2}\\
    0 & 0 & 0 & \lambda_{LRN} y^\nu_{33} \frac{v_u^2}{M_R} & \lambda_{3 L} v_{\bar \Phi} \\
    0 & 0 & \lambda_{LRN} y^\nu_{33} \frac{v_u^2}{M_R} & 0 & \lambda_L v_\Phi \\
   \lambda_{2L_1}  &  \lambda_{2L_2}  & \lambda_{3 L} v_{\bar \Phi} & \lambda_L v_\Phi & \lambda_{RLN}^2 \frac{v_d^2}{M_L} \end{array} \right)
   \left( \begin{array}{c} \nu_L^1 \\ \nu_L^2 \\ \nu_L^3 \\ N_L^\prime \\ \bar N_L^{\prime} \end{array} \right) } \end{equation}
The set of parameters for the two fit points has been given in \hyperref[sec:7.5.2]{Appendix~\ref*{sec:7.5.2}} and \hyperref[sec:7.8.2]{Appendix~\ref*{sec:7.8.2}}. \\
The neutrino mass matrix $M_\nu$ can be diagonalized as follows:
\begin{equation}
    U_\nu^T M_\nu U_\nu = diag(m_{\nu_1},m_{\nu_3},m_{\nu_3},m_{\nu_4},m_{\nu_5})
\end{equation}
From this, one can obtain the extended PMNS matrix given by:
\begin{equation}
    U_{PMNS} = (U_L^e)^\dag P_{\bar{5}} U_\nu
\end{equation}
As can be clearly seen from the above definition, the extended PMNS matrix is not unitary. The 3 $\times$ 3 submatrix of this extended PMNS matrix is the standard PMNS matrix. This is the same as saying that, the PMNS matrix is the matrix that diagonalizes the neutrino mass matrix when it is represented in the  basis in which the charged lepton mass matrix is diagonal. The three mixing angles ($sin^2\theta_{12}, sin^2\theta_{23}, sin^2\theta_{13}$) can then be extracted from the PMNS matrix. The light neutrino masses are given by $m_i = \sqrt{m_{\nu_i}^* m_{\nu_i}}$ (i = 1,2,3) and we get 2 neutrinos with mass of order 1 TeV. From the masses of the three light neutrinos we can get the three mass differences $\Delta m^2_{12}, \Delta m^2_{23}$ and $\Delta m^2_{13}$. We also get the $CP$ violating phase $\delta_{CP}$ and Jarlskog invariant defined by: 
\begin{equation}
    J = \textrm{Imag}\left(U_{PMNS}^{23} U_{PMNS}^{13*} U_{PMNS}^{12} U_{PMNS}^{22*} \right)
\end{equation}
which is consistent with the $3\sigma$ confidence interval for $\delta_{CP}$ found in \cite{T2K:2019bcf}.

\subsection{RG flow of the $U(1)_{3-4}$  coupling $g'$}
We consider two loop RG equations for the gauge couplings and 1 loop RG equations for the Yukawa parameters and for the soft SUSY breaking dimensionful parameters - the sparticle mass parameters, gaugino mass paramaters and the parameter $h^{ijk}$. \\
We have an additional constraint on the value of the $g'$ coupling due to the presence of a Landau pole at high energies. The RGE of the coupling $g'$ to 1-loop is given by:
\begin{equation}
    \frac{dg'}{d ln\mu} = \frac{g'^3}{16\pi^2} (\sum_i Q^\prime(i) - 3C(G))
\end{equation}
Here $C(G) = 0$ in our case, since $G = U(1)_{3-4} \equiv U(1)^\prime$.  $Q^\prime(i)$ is the $U(1)^\prime$ charge. The sum is over all the supermultiplets which are charged under the $U(1)_{3-4}$ group. Hence, we get:
\begin{equation}
    \frac{dg'}{d ln\mu} = 34 \frac{g'^3}{16\pi^2}
\end{equation}
This implies:
\begin{equation}
    \Lambda_{g'} = \mu_{Z'} exp(\frac{8\pi^2}{34 g'(\mu_{Z'})^2})
\end{equation}
where $\Lambda_{g'}$ is the scale of the Landau pole. The plot for the scale of the Landau pole as a function of $g'( \sim 1$ TeV) is given in Figure 1 above.
\begin{figure}[]
    \centering
    \includegraphics[width=8cm]{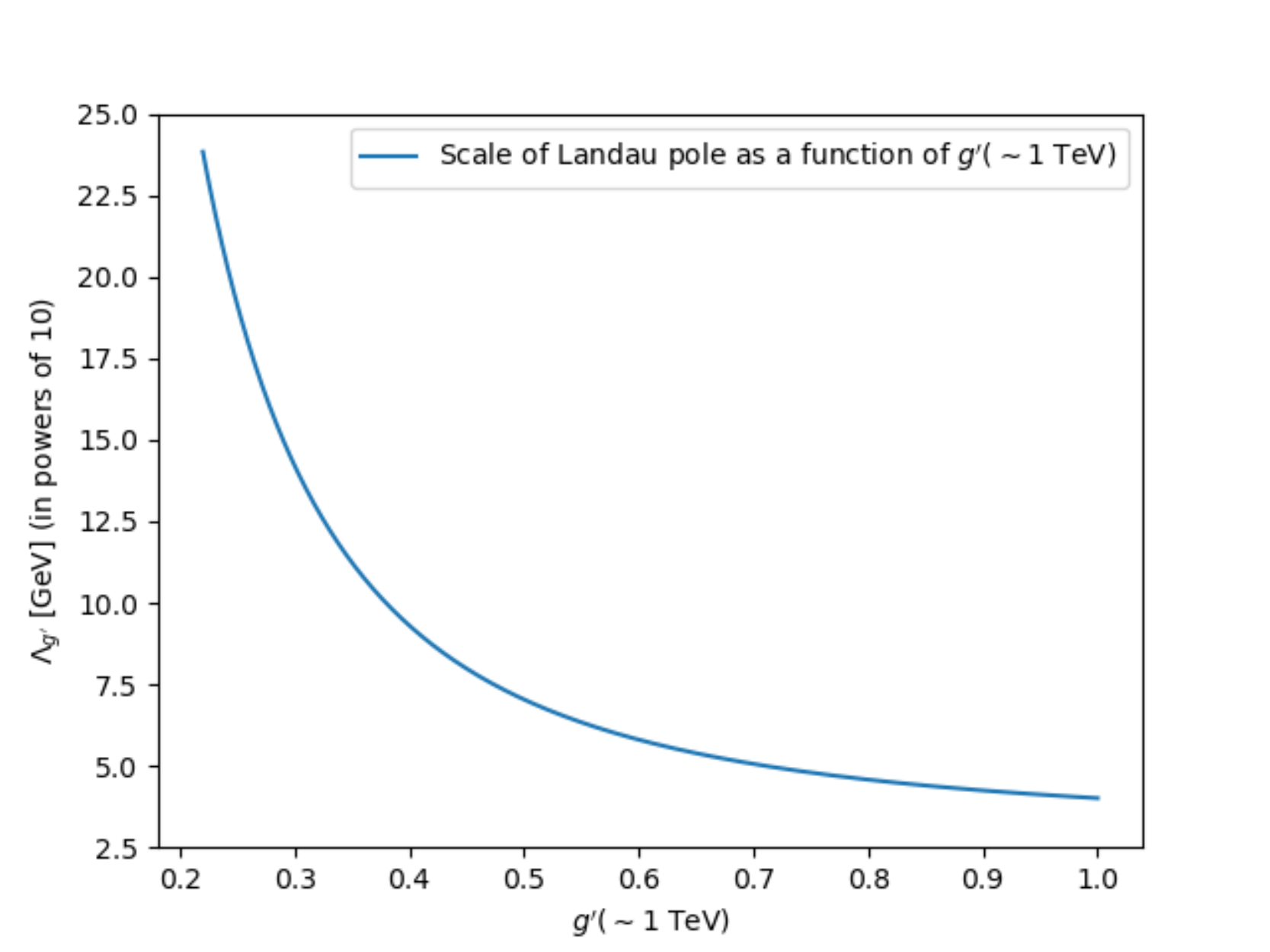}
    \captionsetup{justification=centering}
    \caption{Scale of Landau pole as a function of $g'(\sim 1$ TeV)}
    \label{fig:my_label}
\end{figure}
For our case, we consider our model to hold true roughly up to $M_{GUT} \approx 5 \times 10^{16}$ GeV and consider $\mu(Z') \sim$ TeV scale, within reach of future experiments at the HL-LHC. This already gives a constraint $g'(\mu_{Z'}) < 0.271$ for the $U(1)_{3-4}$ coupling in our model. It should be noted that this is an estimate obtained only from the 1 loop RG equations and not considering the 2 loop effects or the 1 loop threshold corrections, which is what we will be considering for a more detailed analysis.

\subsection{Couplings of quarks and charged leptons with $Z'$ }
The Lagrangian for the coupling of the quarks and charged leptons with the new $Z'$ boson is given by:
\begin{equation}
    \mathcal{L}_{Z'} = g'Z'_{\mu}(\Hat{\Bar{f}}^a_L [\Hat{g}^f_L]_{ab}\Hat{f}^b_L + \Hat{\Bar{f}}^a_R [\Hat{g}^f_R]_{ab}\Hat{f}^b_R)
\end{equation}
where $f = u,d,e$ and
\begin{equation}
    \Hat{g}^f_L = (U^f_L)^\dag g_L U^f_L, \ \Hat{g}^f_R = (U^f_R)^\dag g_R U^f_R
\end{equation}
with the $U(1)_{3-4}$ charge matrices given by:
\begin{equation}
    g_L = g_R = diag(0,0,1,-1,0)
\end{equation}
In a completely analogous manner, the $Z'$ gaugino similarly couples to the quarks, charged leptons and their sparticles. Since the structure of the Lagrangian is not different for them, we won't bother writing it down explicitly. It is worth noting that the $Z'$ coupling is not symmetric in the $4$ and the $\bar 4$ families because of the difference in the $U(1)_{3-4}$ charges in the $4$ and $\bar 4$ families and the difference in the way the $f_R$ and $f_L$ are represented. The flavor-changing currents between the SM 1 and 2 families and the SM 2 and 3 families mediated by the $Z'$ boson are small, leading to suppressed branching ratios for lepton flavor violating processes. This will be explicitly seen in the results of our model's analysis.

\section{OBSERVABLES}
\subsection{Masses of the particles}
We fit the SM quark and charged lepton masses by diagonalizing the 5 $\times$ 5 Dirac mass matrices mentioned in the previous section. The light SM neutrino masses are obtained as mentioned in the previous section. We get that $U_\nu^T M_\nu U_\nu$ is approximately diagonal (with the non-diagonal entries being $\lesssim 0.01$ times the diagonal ones) and the PMNS matrix being approximately unitary.\\
The tree level $W$ and $Z$ boson masses are given by:
\begin{equation}
    M_W = \frac{1}{2}vg_{SU(2)}, \ M_Z = \frac{1}{2}v \sqrt{g_{SU(2)}^2 + g_{U(1)}^2}
\end{equation}
where $v$ is the Higgs VEV, $g_{SU(2)} = g_2$ and $g_{U(1)} = \sqrt{3/5}~ g_1$. However, there are contributions of the new vector-like chiral leptons to the $W$ boson mass, which cause the value of the $W$ boson to deviate from the SM value. $W$ boson mass corrections have been discussed in \cite{Kawamura:2022fhm, deGiorgi:2022xhr, Belfatto:2023tbv}. In our case, we consider the correction to $W$ boson mass due to vector-like chiral leptons given by \cite{Kawamura:2022fhm}:
\begin{equation}
    \frac{\delta M_W^2}{M_W^2|_{SM}} \approx \frac{\alpha_e cos^2(\theta_W)}{cos^2(\theta_W) - sin^2(\theta_W)} T
\end{equation}
where the oblique parameter $T$ is approximately given by:
\begin{equation}
    T \approx \frac{1}{16 \pi sin^2(\theta_W) cos^2(\theta_W)} \frac{2(m_{L^-}^2 - m_{L^0}^2)^2}{3 M_Z^2 M^2_L } 
\end{equation}
with $M^2_L = \lambda_L^2 v_\Phi^2 + \lambda_{2L_1}^2 + \lambda_{2L_2}^2$. The two fit points of our model predict a significant deviation of the $W$ boson mass from its SM value. \\
\\
The Higgs mass is given by:
\begin{equation}
    m^2_{h^0} = \frac{1}{2}(m^2_{A^0} + m^2_Z - \sqrt{(m^2_{A^0}-m_Z^2)^2 + 4m^2_Zm^2_{A^0}sin^2(2\beta)}) + \Delta(m^2_{h^0})
\end{equation}
where $\Delta(m^2_{h^0})$ consists of the radiative corrections to the tree level Higgs mass. The precise expressions are given in detail in the \hyperref[sec:7.2]{Appendix~\ref*{sec:7.2}}. \\
The $Z^\prime$ boson mass in our model is given by:
\begin{equation}
    M_{Z^\prime} = g^\prime \sqrt{2(v_\Phi^2 + v_{\bar \Phi}^2)}
\end{equation}
Its value for the two fit points can be found in \textbf{Table 4}.

\subsection{Masses of the sparticles}
We work in the Generalized Mirage Mediation (GMM) model \cite{Baer:2016hfa} and set the boundary conditions for the sparticle and gaugino masses, and the trilinear soft coupling to be:
\begin{equation}
    M_a = M_s(l_a \alpha + b_a g_a^2)
\end{equation}
\begin{equation}
    A_{ijk} = M_s(-a_{ijk} \alpha + \gamma_i + \gamma_j + \gamma_k)
\end{equation}
\begin{equation}
    m^2_{i} = M_s^2(c_i \alpha^2 + 4\alpha \xi_i - \Dot{\gamma}_i)
\end{equation}
where $M_s = \frac{m_{3/2}}{16\pi^2}$ ($m_{3/2}$ is the gravitino mass), $b_a$ are the gauge $\beta$ function coefficients for the gauge group $a$ and $g_a$ are the corresponding gauge couplings and $\xi_i = \sum_{j,k} a_{ijk}\frac{y_{ijk^2}}{4} - \sum_a l_a g_a^2 C^a_2(f_i)$ with $C^a_2(f_i)$ being the quadratic Casimir for the $a$'th gauge group corresponding to the representation to which the sfermion $\Tilde{f}_i$ belongs. The anomalous dimension $\gamma_i$ and $\Dot{\gamma}_i = 8\pi^2\frac{\partial \gamma_i}{\partial log\mu}$ is given by:
\begin{equation}
    \gamma_i = 2\sum_{a} g_a^2 C^a_2(f_i) - \sum_{y_i} |y_i|^2
\end{equation}
\begin{equation}
    \Dot{\gamma}_i = 2\sum_{a} g_a^4 b_a C^a_2(f_i) - \sum_{y_i} |y_i|^2 b_{y_i}
\end{equation}
We consider a universal parameter $a_{ijk} = a_0$ and a universal $c_i = c_m$ for all the sparticles except for $m^2_{H_u}$ and $m^2_{H_d}$ for which we consider the coefficients $c_{H_u}$ and $c_{H_d}$. Hence the input parameters for these soft SUSY breaking sector include $M_s, l_a, \alpha, a_0, c_m, c_{H_u}$ and $c_{H_d}$. The values for these input parameters at the GUT scale are mentioned in \hyperref[sec:7.5.3]{Appendix~\ref*{sec:7.5.3}} and \hyperref[sec:7.8.3]{Appendix~\ref*{sec:7.8.3}}.
The sparticle masses predicted by the model are all less than 25 TeV. The gluino mass is $\sim ~ 2.3 - 3$ TeV and satisfies the bound $> 1.9$ TeV. The VL particle and the $Z'$ masses are all of the $\sim$ TeV scale within reach of future experiments. The neutralino and chargino masses are found by diagonalizing the $7 \times 7$ and $2 \times 2$ matrices given in the \hyperref[sec:7.3]{Appendix~\ref*{sec:7.3}}.

\subsection{Lepton Non-Universality}
Our model gives rise to Lepton Non-Universality in the operators $\mathcal{O}_{i = 9,10}$ and $\mathcal{O}'_{i = 9,10}$. The effective contributions to the Wilson coefficients are given by:
\begin{equation}
    C_i = -{g^\prime}^2\frac{\sqrt{2}}{G_F}\frac{1}{V_{tb}V_{ts}^*}\frac{\pi}{\alpha_e}\frac{1}{2M_{Z^\prime}^2} g_{eff,i}
\end{equation}
and
\begin{equation}
    C'_i = -{g^\prime}^2\frac{\sqrt{2}}{G_F}\frac{1}{V_{tb}V_{ts}^*}\frac{\pi}{\alpha_e}\frac{1}{2M_{Z^\prime}^2} g'_{eff,i}
\end{equation}
with the couplings:
\begin{equation}
    g_{eff,9} = [\Hat{g}^d_L]_{23}[\Hat{g}^e_R + \Hat{g}^e_L]_{22}, \ g_{eff,10} = [\Hat{g}^d_L]_{23}[\Hat{g}^e_R - \Hat{g}^e_L]_{22}
\end{equation}
and
\begin{equation}
    g'_{eff,9} = [\Hat{g}^d_R]_{23}[\Hat{g}^e_R + \Hat{g}^e_L]_{22}, \ g'_{eff,10} = [\Hat{g}^d_R]_{23}[\Hat{g}^e_R - \Hat{g}^e_L]_{22}
\end{equation}
Our analysis presented here is only concerned with all the rare $B$ decays i.e. the $b \longrightarrow s \mu \mu$ observables, LFU observables and the $B_s \longrightarrow \mu^+ \mu^-$ branching ratio. The relevant Wilson coefficients have been calculated by using the $\Hat{g}$ matrices in \hyperref[sec:7.7]{Appendix~\ref*{sec:7.7}} and \hyperref[sec:7.10]{Appendix~\ref*{sec:7.10}} for the two fit points I and II respectively. Recent results \cite{LHCb:2022zom} suggest that the rare $B$ decay anomalies found earlier are not present and hence in this scenario, we must fit the Wilson coefficients to C$^\mu_9$, C$^\mu_{10}$, C${^\prime}^\mu_9$ and C${^\prime}^\mu_{10}$ to approximately zero. This is realized in our model where we fit them to $0.0 \pm 0.1$.

\subsection{Electric dipole moment of the muon}
Following \cite{Hamaguchi:2022byw}, the muon electric dipole moment is given by:
\begin{equation}
    d_\mu = d^h_\mu + d^Z_\mu + d^W_\mu
\end{equation}
where
\begin{equation}
    d^h_\mu = - \frac{\sqrt{4\pi \alpha_e}}{16\pi^2 M_{Higgs}^2} \sum_{i = 2,4,5} Im(\lambda_{i2} \lambda_{2i}) m_i g_h(r_i)
\end{equation}
where $r_i = m_i^2/M_{Higgs}^2$ and $\lambda_{ij}$ are the Yukawa couplings between the charged leptons $f_i$ and $f_j$. Since $\lambda_{i2}$ and $\lambda_{2i}$ are real for both our fit points, $d^h_\mu = 0$. $d^Z_\mu$ and $d^W_\mu$ are given by:
\begin{equation}
    d_\mu^Z = \frac{\sqrt{4\pi \alpha_e}}{16\pi^2 M_{Z}^2} \sum_{i=2,4,5} Im\left([g_L^Z]_{i2} [g_R^{Z*}]_{i2}\right) m_i g_Z(r_i)
\end{equation}
where $g_L^Z$ and $g_R^Z$ are the left and right handed couplings of charged leptons to the Z boson.
\begin{equation}
    d_\mu^W = \frac{\sqrt{4\pi \alpha_e}}{16\pi^2 M_{W}^2}  Im\left([g_L^W]_{42} [g_R^{W*}]_{42}\right) m_{\nu_4} g_W(r_4)
\end{equation}
Here $r_4 = m_{\nu_4}^2/M_W^2$, and $g_L^W$ and $g_R^W$ are the left and right handed couplings of charged leptons to the W boson. The functions $g_h(r_i)$, $g_h(r_i)$ and $g_h(r_i)$ can be found in \cite{Hamaguchi:2022byw}. The muon electric dipole moment in our model is consistent with the latest experimental data.

\subsection{Electric dipole moment of the electron}
Again following \cite{Hamaguchi:2022byw}, the electron electric dipole moment is given by:
\begin{equation}
    d_e = d_e^{hV\gamma} + d_e^{WW\gamma}
\end{equation}
where $V = \gamma, Z$ and
\begin{equation}
    d_e^{hV\gamma} = \frac{\sqrt{4\pi \alpha_e} g_\nu^{Vee} g_s^{ee}}{32 \pi^4} \sum_{i,j = 2,4,5} \int_0^1 dx Im(c_O^{Vij}) I^{ij}_{hV}
\end{equation}
where $g_\nu^{Vee}$ and $g_s^{ee}$ are the SM couplings of the electron to vector bosons $V = \gamma, Z$ and Higgs boson. Their precise expressions and those of other functions appearing in the above equation can be found in \cite{Hamaguchi:2022byw}. In our model $d_e^{hV\gamma}$ is zero because $c_O^{Vij}$ is purely real because the only complex charged lepton Yukawa coupling in our model is $y_\tau$. \\
The expression of $d_e^{WW\gamma}$ is given by:
\begin{equation}
    d_e^{WW\gamma} = \frac{e g^2}{256 \pi^4} \sum_{j = 4,5} \sum_{i = 1,2,3,4,5} \frac{m_e m_i m_{\nu_j}}{M_W^2} Im([g_L^W]_{ji} [g_R^{W*}]_{ji}) \int_0^1 dx I^{ji}_{WW}
\end{equation}
where
\begin{equation}
    I^{ji}_{WW}(x) = \frac{(1-x)}{-x(1-x)M_W^2 + (1-x)m^2_i + x m_{\nu_j}^2} ln\frac{(1-x)m^2_i + x m_{\nu_j}^2}{x(1-x)M_W^2}
\end{equation}
with $j \in \{4,5\}$. We find that the electric dipole moment of the electron $d_e$ for the two fit points of our model is consistent with the latest experimental bounds.

\subsection{Anomalous magnetic moment of the muon}
There are discrepancies between experiments and the SM prediction for the magnetic moment of the muon:
\begin{equation}
    \Delta a_\mu = a_\mu^{exp} - a_\mu^{SM} = 251(59) \times 10^{-11}
\end{equation}
The $W$ and  $Z$ contributions to the anomalous magnetic moment of the muon are very close to their SM values and hence we do not consider them here. Since we are working in the large $tan\beta$ regime, there might be sizeable contributions to the anomalous magnetic moment of the muon from the Higgs. Moreover, there are SUSY contributions to the muon $g - 2$ as well and the $Z'$ contribution can be sizeable due to the off-diagonal muon-$Z'$ couplings to the VL leptons which contribute significantly to the muon $g$ - 2 loop.

The leading order contribution to the muon $g$ - 2 is then given by:
\begin{equation}
    \Delta a_\mu = \Delta a_\mu^{Z'} + \Delta a_\mu^{Higgs} + \Delta a_\mu^{SUSY}
\end{equation}
where the leading order contribution to the muon $g$ - 2 arising from the $Z'$ coupling to leptons is given by (for example \cite{Dermisek:2013gta,Jegerlehner:2009ry}):
\begin{equation}
    \Delta a_\mu^{Z'} = - {g^{\prime}}^2\frac{m_\mu^2}{8 \pi^2 M_{Z'}^2} \sum_{a}\left[(|[\Hat{g}^e_L]_{2a}|^2 + |[\Hat{g}^e_R]_{2a}|^2)F(x_a) + Re([\Hat{g}^e_L]_{2a}[\Hat{g}^{e*}_R]_{2a}) \frac{m_a}{m_\mu}G(x_a)\right]
\end{equation}
where $x_a = (m_a/M_{Z'})^2$ and
\begin{equation}
    F(x) = (5x^4 - 14x^3 + 39x^2 - 38x - 18x^2ln(x) + 8)/(12 (1 - x)^4)
\end{equation}
\begin{equation}
    G(x) = (x^3 + 3x - 6xln(x) - 4)/(2 (1 - x)^3)
\end{equation}
and the Higgs contribution is given by \cite{Dermisek:2022hgh}:
\begin{equation}
    \Delta a_\mu^{Higgs} \approx 2\frac{1+tan^2\beta}{16\pi^2} \frac{m_\mu m_\mu^{LE}}{v^2}
\end{equation}
where
\begin{equation}
    m_\mu^{LE} = \frac{\lambda_{2L_2} \lambda_{2E_2} \lambda_{RLE}}{m_{E_1}m_{E_2}} v_{H_u}
\end{equation}
and the SUSY contributions to the muon $g - 2$ are as given in \cite{Athron:2021iuf}. In our analysis, we have found fit points with $\lambda_{2E_2} = 0$ and hence, this gives a vanishing value of $\Delta a_\mu^{Higgs}$. The $\Delta a_\mu^{SUSY}$ in our model are also smaller than the $\Delta a_\mu^{Z'}$ and the muon $g - 2$ is hence dominated by $\Delta a_\mu^{Z'}$. It is worth mentioning that one might also solve the muon $g - 2$ anomaly by making $\lambda_{2E_2} \sim \lambda_{2L_2}$ but keeping $\lambda_{RLE}$ small enough (of the order of $10^{-6}$). In such scenarios, the muon $g - 2$ is dominated by $\Delta a_\mu^{Higgs}$ and the contribution of $\Delta a_\mu^{Z'}$ is much smaller in these cases.

\subsection{Lepton Flavor violating processes}
The new $Z'$ coupling induces the Lepton flavor violating decays $\mu \longrightarrow e \gamma$, $\tau \longrightarrow \mu \gamma$ and $\tau \longrightarrow \mu \mu \mu$. Apart, from this, there are also dominant SUSY contributions to $BR(\mu \longrightarrow e \gamma)$ and $BR(\tau \longrightarrow \mu \gamma)$. There is a constraint on the branching ratios of these decays due to the latest experimental measurements giving rise to upper bounds on these branching ratios. \\ Using the general expressions in \cite{Lavoura:2003xp}, we can calculate the branching ratios of $\mu \longrightarrow e \gamma$ and $\tau \longrightarrow \mu \gamma$. (See also \cite{Hisano:1995cp,Ishiwata:2013gma,Abada:2014kba})\\
The branching ratio for the process $\mu \longrightarrow e \gamma$ is given by:
\begin{equation}
    BR(\mu \longrightarrow e \gamma) \approx \frac{1}{\Gamma_\mu} \frac{\alpha_e m_\mu^3}{1024 \pi^4}\frac{{g^\prime}^2}{2 M_{Z^\prime}^2}\left(|\sum_{a \in VL}[\Hat{g}^e_L]_{1a}[\Hat{g}^e_R]_{a2}|^2 + |\sum_{a \in VL}[\Hat{g}^e_R]_{1a}[\Hat{g}^e_L]_{a2}|^2\right) + BR(\mu \longrightarrow e \gamma)^{SUSY}
\end{equation}
The branching ratio for the process $\tau \longrightarrow \mu \gamma$ is given by:
\begin{equation}
    BR(\tau \longrightarrow \mu \gamma) \approx \frac{1}{\Gamma_\tau} \frac{\alpha_e m_\tau^3}{1024 \pi^4}\frac{{g^\prime}^2}{2 M_{Z^\prime}^2}\left(|\sum_{a \in VL}[\Hat{g}^e_L]_{2a}[\Hat{g}^e_R]_{a3}|^2 + |\sum_{a \in VL}[\Hat{g}^e_R]_{2a}[\Hat{g}^e_L]_{a3}|^2\right) + BR(\tau \longrightarrow \mu \gamma)^{SUSY}
\end{equation}
where the expressions for $BR(\mu \longrightarrow e \gamma)^{SUSY}$ and $BR(\tau \longrightarrow \mu \gamma)^{SUSY}$ have been calculated using the equations (4.11) and (4.12) of \cite{Dedes:2007ef}.
For the branching ratio of $\tau \longrightarrow \mu \mu \mu$, we follow the reference \cite{Okada:1999zk,Kuno:1999jp} to get:
\begin{equation}
    BR(\tau \longrightarrow \mu \mu \mu) \approx \frac{1}{\Gamma_\tau} \frac{m_\tau^3}{1536 \pi^3}\frac{{g^\prime}^4}{M_{Z^\prime}^4}\left(2|[\Hat{g}^e_L]_{22}[\Hat{g}^e_L]_{23}|^2 + 2|[\Hat{g}^e_R]_{22}[\Hat{g}^e_R]_{23}|^2 + |[\Hat{g}^e_L]_{22}[\Hat{g}^e_R]_{23}|^2 + |[\Hat{g}^e_R]_{22}[\Hat{g}^e_L]_{23}|^2\right)
\end{equation}
Our model predicts these branching ratios for these processes within the known experimental bounds. The fit points we have obtained give $BR(\mu \longrightarrow e \gamma)$ within reach of future experiments. One of the fit points have a $BR(\mu \longrightarrow e \gamma)$ that is reasonably close to the current experimental bound while the other fit point has a relatively smaller value of $BR(\mu \longrightarrow e \gamma)$.

\subsection{Quark Flavor changing processes}
We will discuss three quark flavor changing processes, namely $B_s - \bar{B}_s$, $B_s \longrightarrow \mu^+ \mu^-$, and $K_L \longrightarrow \mu^+ \mu^-$ and show that the model predictions at the two fit points are consistent with latest experimental data.
\subsubsection{$B_s - \bar{B}_s$ mixing}
Adopting the numerical factors mentioned in \cite{Altmannshofer:2014cfa,Buras:2012jb} the relative change in the mixing matrix element due to the tree level $Z^\prime$ mixing contribution to $B_s \longleftrightarrow \bar{B}_s$ is given by:
\begin{equation}
    \delta M_{12} \approx \left(2.3 \frac{g^2}{16\pi^2 M_W^2} (V_{ts} V_{tb})^2\right)^{-1} \frac{{g^\prime}^2}{M_{Z^\prime}^2} (|[\hat{g}^d_L]_{32}|^2 + |[\hat{g}^d_R]_{32}|^2 + 9.7Re([\hat{g}^d_L]_{32}[\hat{g}^{d,*}_R]_{32}))
\end{equation}
Both the fit points of our model predict a deviation of $\delta M_{12} < \pm 6\%$ from the SM which is consistent with the most recent theoretical uncertainty and experimental data.
\subsubsection{$B_s \longrightarrow \mu^+ \mu^-$}
The new particles give rise to new physics contributions to the process $B_s \longrightarrow \mu^+ \mu^-$. The effective Lagrangian relevant for our analysis is of the form:
\begin{equation}
    -\mathcal{L}^{NP}_{eff} = (\bar s \gamma_\mu \gamma_5 b)([\hat{g}^d_L]_{23} - [\hat{g}^d_R]_{23}) \frac{{g^\prime}^2}{4 M_{Z^\prime}^2} \left((\bar \mu \gamma^\mu \mu)([\hat{g}^e_L]_{22} + [\hat{g}^e_R]_{22}) + (\bar \mu \gamma^\mu \gamma_5 \mu)([\hat{g}^e_R]_{22} - [\hat{g}^e_L]_{22})\right)
\end{equation}
Hence, in this model, the coefficient $C^V_{AA}$ is given by:
\begin{equation}
    C^V_{AA} = C_{SM}(B_s) + \frac{{g^\prime}^2}{4 M_{Z^\prime}^2} ([\hat{g}^d_L]_{23} - [\hat{g}^d_R]_{23})([\hat{g}^e_R]_{22} - [\hat{g}^e_L]_{22})
\end{equation}
where
\begin{equation}
    C_{SM}(B_s) = \frac{4 G_F}{\sqrt{2}} \frac{\alpha_e}{2\pi sin^2\theta_W} V_{ts}^* V_{tb} \eta_Y Y_o(x_t)
\end{equation}
and $\eta_Y = 1.012$ quantifies the QCD corrections and $Y_0(x_t)$ is the loop function given by:
\begin{equation}
    Y_0(x_t) = \frac{x_t}{8}\left(\frac{x_t-4}{x_t-1} + \frac{3x_t}{(x_t - 1)^2}log(x_t)\right)
\end{equation}
with $x_t = \frac{m_t^2}{M_W^2}$. \\
The ratios of the branching fractions of our model and the SM are then given by:
\begin{equation}
    R^{th}_{B_s \longrightarrow \mu \mu} = \frac{\overline{BR}(B_s \longrightarrow \mu^+ \mu^-)}{\overline{BR}(B_s \longrightarrow \mu^+ \mu^-)_{SM}} = \frac{1 + A_{\Delta \Gamma}y_s}{1 + y_s} |P_s|^2
\end{equation}
Here, $P_s = |P_s| e^{i\phi_P} = \frac{C^V_{AA}}{C_{SM}(B_s)}$, $y_s = 0.065$ and $A_{\Delta \Gamma} = cos(2\phi_P - \theta_S^{B_s})$. To evaluate the phase $\theta_S^{B_s}$, we follow \cite{Buras:2012jb}. Hence, we have:
\begin{equation}
    S(B_s) = S_0(x_t) + \Delta S(B_s) = |S(B_s)| e^{i\theta_S^{B_s}}
\end{equation}
where
\begin{equation}
    S_0(x_t) = \frac{4x_t - 11x_t^2 + x_t^3}{4(1-x_t)^2} - \frac{3x_t^2 logx_t}{2(1-x_t)^3}
\end{equation}
with $x_t = \frac{m_t^2}{M_W^2}$ and
\begin{equation}
    \Delta S(B_s) = \left(\frac{[\hat{g}^d_L]_{32}}{V_{tb}^* V_{ts}}\right)^2 \frac{4 \Tilde{r}}{M_{Z^\prime}^2 g_{SM}^2} + \left(\frac{[\hat{g}^d_R]_{32}}{V_{tb}^* V_{ts}}\right)^2 \frac{4 \Tilde{r}}{M_{Z^\prime}^2 g_{SM}^2} + \Delta S(B_s)_{LR}
\end{equation}
where $\Tilde{r} = 0.985$, $g_{SM}^2 = \frac{4 G_F}{\sqrt{2}} \frac{\alpha_e}{2\pi sin^2\theta_W}$ and $\Delta S(B_s)_{LR}$ is evaluated using the equations (23) and (25) and using the relevant values in Table 1 and 3 of \cite{Buras:2012jb}. \\
From the experimental values of $\bar{BR}(B_s \longrightarrow \mu^+ \mu^-)$ \cite{ParticleDataGroup:2018ovx} and the SM predictions \cite{Altmannshofer:2017wqy, Bobeth:2013uxa}, it is known that:
\begin{equation}
    R^{exp}_{B_s \longrightarrow \mu \mu} = \frac{\overline{BR}(B_s \longrightarrow \mu^+ \mu^-)_{exp}}{\overline{BR}(B_s \longrightarrow \mu^+ \mu^-)_{SM}} = 0.75 \pm 0.16
\end{equation}
We find that the two fit points in our model predict $R^{th}_{B_s \longrightarrow \mu \mu}$ that is consistent with the above experimental evidence. We note that this is the only dominant contribution in our model, because a possible enhancement due to large tan$\beta$ as indicated in \cite{Buras:2002wq} is not significant here due to $m^2_{A_0} = 2|\mu|^2 + m_{H_u}^2 + m_{H_d}^2$ being large ($ \sim \mathcal{O}(100)$ TeV$^2$) in our model, and due to the process being suppressed by $\frac{1}{m_{A_0}^4}$. 

\subsubsection{$K_L \longrightarrow \mu^+ \mu^-$}
Following \cite{Buras:2004ub}, we get $BR(K_L \longrightarrow \mu^+ \mu^-)$ to be:
\begin{equation}
    BR(K_L \longrightarrow \mu^+ \mu^-) = 2.08 \times 10^{-9}\left[\Bar{P_c}(Y_K) + \frac{|V_{cb}|^2}{\lambda^2} R_t |Y_A(K)| cos(\Bar{\beta}^K_Y)\right]^2
\end{equation}
where $\lambda = 0.226$, $\Bar{\beta}^K_Y = \beta - \beta_s - \theta^K_Y$ ($V_{td} = |V_{td}|e^{-i\beta}$, $V_{ts} = -|V_{ts}|e^{-i\beta_s}$), $\theta^K_Y$ is the phase associated with $Y_A(K)$. $\Bar{P_c}(Y_K)$ is given by:
\begin{equation}
    \Bar{P_c}(Y_K) = 0.113 \times (1 - \frac{\lambda^2}{2})
\end{equation}
and the expressions of $R_t$ and $Y_A(K)$ are as mentioned below:
\begin{equation}
    R_t = \frac{|V_{td}|}{|V_{us}||V_{cb}|}
\end{equation}
and
\begin{equation}
    Y_A(K) = |Y_A(K)| e^{i\theta_Y^K} = \eta_Y Y_0(x_t) + \frac{([\hat{g}^e_R]_{22} - [\hat{g}^e_L]_{22})}{M_{Z^\prime}^2 g_{SM}^2}\frac{([\hat{g}^d_L]_{21} - [\hat{g}^d_R]_{21})}{V_{ts}^* V_{td}}
\end{equation}
The value for $BR(K_L \longrightarrow \mu^+ \mu^-)$ predicted by our model for the two fit points is consistent with the experimental bound of $BR(K_L \longrightarrow \mu^+ \mu^-) \leq 2.5 \times 10^{-9}$.

\section{METHODOLOGY}
We employ a mix of guesswork and $\chi^2$ minimization technique in \texttt{python} to find the fit points in the parameter space. The analysis has been performed by considering the two-loop RG equations for the gauge couplings and their one-loop threshold corrections and one-loop RG equations for the Yukawa parameters and the soft SUSY breaking parameters by analytically deriving the RG equations for this supersymmetric "VL-chiral" model from the most general expressions in \cite{Martin:1993zk}. We start with the 1-loop RG flow of the gauge couplings and find the value of the unified coupling $\alpha_{unif}$ at $\approx 5 \times 10^{16}$ GeV and then choose the mirage mediation boundary condition parameters and some initial guess for the Yukawa parameters and RG flow these parameters down to the weak scale via 2-loop RG equations for the gauge couplings and 1-loop RG equations for all the other parameters. The coupled differential RG equations are solved using \texttt{scipy.integrate.odeint}. We also require precise
gauge coupling unification with $\epsilon_3 = 0$.

With the weak scale results at hand, the value of $\alpha_{GUT}$ at the GUT scale and the GMM model parameters are then varied to get weak scale gauge couplings that are approximately close to the experimental values. Then the GUT scale Yukawa parameters are varied to fit the low energy observables approximately.

Following that, a $\chi^2$ function is constructed with the GUT scale parameters taken to be those found out from the previous analysis, which are known to give approximately good results at the weak scale. Then \texttt{scipy.optimize.minimize} is used to minimize the $\chi^2$ function and due to limited computing power, the code was terminated when the results were seen to saturate to particular values. The resulting input array then gives the set of GUT scale parameters of the theory. For the neutrino sector analysis, since we have integrated out the massive states at a mass scale of $\approx 10^{12}$ GeV, from this point to the weak scale, we RG flow the Weinberg operator along with the other parameters by using the expressions analogous to the one in \cite{Babu:1993qv, Antusch:2001ck, Antusch:2001vn} from which we extract the neutrino mixing angles and mass differences and include them in the $\chi^2$ minimization procedure. It should be noted that all the couplings associated with the neutrino fields (which are integrated out), have been set to zero while RG flowing the other parameters from the Majorana mass scale to the weak scale. For example, the yukawa couplings $y^\nu, \lambda_{LRN}, \lambda_{RLN}, \lambda_N$ and $\lambda_{3N}$ are all set to zero. Hence the coupled differential RG equations are solved twice with appropriate initial conditions, once as we go down from the GUT scale to the Majorana scale and then again as we go down from the Majorana scale to the weak scale. It is also worth mentioning that the neutrino sector observables (mass differences and mixing angles) can be entirely controlled by changing the neutrino sector parameters: especially the parameters in the matrix $y^\nu_{ij}$, $\lambda_{LRN}$ and the mass parameters that go into the Majorana Lagrangian. Moreover, since these above-mentioned Yukawa parameters are not exceedingly large, changing them by reasonable amounts (roughly by 1 \%, say) does not affect the RG flow of the other Yukawa parameters much and hence keeps the other low energy observables unchanged. Hence, we expect that the neutrino sector low energy observables can be easily controlled by varying only these parameters and they are sufficient to explain the low energy neutrino sector observables without affecting the other observables significantly.

We find two fit points that fit the low energy observables and reasonably different $BR(\mu \longrightarrow e \gamma)$. Hence, we simultaneously fit the CKM matrix, masses of the charged leptons and quarks at the weak scale, neutrino mixing angles, their mass differences, the strong and U(1)$_{em}$ gauge couplings, and the Higgs mass. We solve for the muon $g - 2$ and also see that the $W$ boson mass deviates from its SM value, without violating the experimental bounds of the electric dipole moments, the LFV processes $\mu \longrightarrow e \gamma$, $\tau \longrightarrow \mu \gamma$, $\tau \longrightarrow \mu \mu \mu$, the quark flavor changing processes $B_s \longrightarrow \mu^+ \mu^-$, $K_L \longrightarrow \mu^+ \mu^-$, $B_s \longleftrightarrow \bar{B}_s$ and the upper bound of $0.12$ eV for $\sum_{i=1}^{3} m_{\nu_i}$ obtained from cosmological observations.

The input parameters that are being varied in our $\chi^2$ analysis can be found in \hyperref[sec:7.5]{Appendix~\ref*{sec:7.5}} and \hyperref[sec:7.8]{Appendix~\ref*{sec:7.8}}, and the observables that are being fitted can be found in \textbf{Table 2} and \textbf{Table 3}. Since there are more input parameters than there are observables to fit, the possibility of another point in parameter space which can fit the low energy observables with different predictions is not excluded.

\section{RESULTS}
\subsection{Low energy observables}

All the gauge couplings and the Yukawa parameters satisfy $g < 1.5$ and $\alpha_y = \frac{Y^2}{4\pi} \lesssim 0.7$ from the weak scale up to the GUT scale and hence the theory is perturbative up to the GUT scale. At the GUT scale, we have also imposed $y_t = y_b = y_\tau$ consistent with grand unification at the GUT scale.\footnote{$y_{\nu_\tau}$ would also be equal in SO(10).} An interesting observation which can be made from the values of these couplings at the weak scale (See \hyperref[sec:7.5]{Appendix~\ref*{sec:7.5}} and \hyperref[sec:7.8]{Appendix~\ref*{sec:7.8}}) is that although the values of these couplings have a different magnitude, the phases associated with these three parameters are the same, even at the weak scale.

We have found two points in the parameter space which give very good fits for the CKM matrix, quark, lepton, Higgs mass, neutrino mass differences, mixing angles and the strong and U(1)$_{em}$ gauge couplings. From the results of the mass squared differences, we can also see that the model favors normal ordering over inverted ordering for neutrino masses. The quark and lepton masses of the three SM families have been fit by following the precise values of the running mass parameters obtained at the $M_Z$ scale as in \cite{Huang:2020hdv}. The total number of input parameters in the model is 63 and the total number of fitted low-energy observables is 46.  It might be possible to reduce the number of input parameters with some symmetry relations.   This is however outside the scope of the present paper.

The Wilson coefficients relevant for lepton non-universality in B decays are fit to $\approx 0$ consistent with \cite{LHCb:2022zom}. It is worth mentioning that the Wilson coefficients are sensitive to the coupling $\lambda_{3Q}$ in a way that changing this parameter slightly, doesn't affect the CKM matrices or weak scale masses much, however, it leads to a change in the value of C$^\mu_9$ that might deviate significantly from the  value we are interested in. The reality of the Wilson coefficients can also be controlled reasonably by the real and imaginary parts of the Yukawa parameter $\lambda_{3Q}$ which determines the coupling of the 3rd SM family quark doublet and the $\bar 4$ quark doublet. The fit points also explain the muon $g - 2$ anomaly and satisfy the experimental bounds of the LFV and quark flavor changing processes, and give the BR($\mu \longrightarrow e \gamma$) within reach of future experiments. Note that the fit point predictions for BR($\mu \longrightarrow e \gamma$) range from $\mathcal{O}$($10^{-15}$) to $\mathcal{O}$($10^{-13}$) and one of these predictions is very close to the present experimental bound on BR($\mu \longrightarrow e \gamma$). It is worth mentioning that BR($\mu \longrightarrow e \gamma$) can be varied easily by changing the sparticle masses, i.e. by changing $m_{3/2}$ and $c_m$ parameters within the GMM model.

The loop contributions to the muon $g$ - 2 from $Z$, $W$, and Higgs are very close to their SM values while the $Z^\prime$ exchange with VL leptons in the loop completely accounts for the anomaly. These good fits can be obtained by considering a point in the parameter space with $\lambda_{3E} = \lambda_{3L} = \lambda_{3N} = 0$ that is, by turning off the interactions between the SM 3rd family leptons with the $\bar 4$ leptons at the GUT scale. At the weak scale as well, these parameters have values of 0 for the two fit points as can be clearly seen from the weak scale Yukawa matrices given in \hyperref[sec:7.5]{Appendix~\ref*{sec:7.5}} and \hyperref[sec:7.8]{Appendix~\ref*{sec:7.8}}. However, it is worth noting that the BR($\tau \longrightarrow \mu \gamma$) and BR($\tau \longrightarrow \mu \mu \mu$) are sensitive to these couplings and changing them to $10^{-6}$, say leads to an enhancement of BR($\tau \longrightarrow \mu \gamma$) and BR($\tau \longrightarrow \mu \mu \mu$) probably within reach of future experiments.

Points in the parameter space with slight perturbations of the input parameters considered will all flow towards the GUT scale parameters for which we have performed the analysis after $\chi^2$ minimization. The Higgs mass is fine-tuned by varying the GMM parameters $c_{H_u}$ and $c_{H_d}$ at the GUT scale and leading order radiative corrections from the top, stop loops, and loops of the VL particles to the Higgs mass have been considered. It is natural to expect the requirement of some fine-tuning for the Higgs mass since the sparticle and gaugino masses are all of the order of 1 - 30 TeV which is necessary if we want to satisfy the gluino mass bound and BR($\mu \longrightarrow e \gamma$) within the Generalized Mirage Mediation model. As a passing note, it is also worth mentioning that scaling $v_\Phi$, $v_{\bar \Phi}$ and the effective $\mu$-like dimension 1 parameters by the same factor keeps the CKM matrix and the SM quark and lepton masses almost unchanged. Also, $\lambda_{2N_1}$ and $\lambda_{2N_2}$ do not appear in our analysis and are hence unconstrained.

The model also predicts masses of the VL leptons and quarks satisfying the experimental bounds and within reach of future experiments. The additional $Z'$ gauge boson and also the two Dirac neutrinos have a TeV scale mass. All the computed and experimental values of all the low energy observables (fitted and unfitted) have been presented in the \textbf{Table 2},  \textbf{Table 3} and \textbf{Table 4}  below. The model also gives $sin^2(\theta_W) = 0.23716$ (for fit point I) and $sin^2(\theta_W) = 0.23740$ (for fit point II) which are around 2.7$\%$ different from the experimental value of 0.23121. It is worth mentioning that the Leading Logarithmic Threshold (LLT) approximation used in calculating the threshold corrections for $\alpha_1$ and $\alpha_2$, which were then used to calculate $sin^2(\theta_W)$, is not consistent since the $SU(2)_L \times U(1)_Y$ symmetry is broken at a scale comparable to the scale at which we have integrated out the masses. Due to this symmetry breaking, we think of our model as the effective theory above the $M_Z$ scale and the $SU(3)_C \times U(1)_{em}$ as the effective theory below the $M_Z$ scale which allows us to consistently fit the gauge couplings $\alpha_3$ and $\alpha_{em}$ and gives us a small theoretical uncertainty $\sim 2.7 \%$ in the calculation of $sin^2(\theta_W)$. 
The $Z$ boson mass is obtained from its tree level expression and is within $1 - 3\%$ of the experimental value. The Higgs mass has been fine tuned within the 1$\sigma$ bound of the experimental value. 

Under RG flow, the $m^2_\Phi$ and $m^2_{\bar \Phi}$ parameters are run down to negative values at the weak scale. However, even in that case, the parameters $m^2_\Phi + |\mu_\Phi|^2$ and $m^2_{\bar \Phi} + |\mu_\Phi|^2$ that appear in the classical potential for the new scalar fields are positive and satisfy the condition of $(b_\Phi \mu_\Phi)^2 > (m^2_\Phi + \mu_\Phi^2)(m^2_{\bar \Phi} + \mu_\Phi^2)$ for U(1)$^\prime$ symmetry breaking. The respective values of $m^2_\Phi$ and $m^2_{\bar \Phi}$ at the weak scale are given by $-211.47$ TeV$^2$ and $-98.745$ TeV$^2$ for the fit point I and $-549.88$ TeV$^2$ and $-336.16$ TeV$^2$ for the fit point II respectively.

\vspace{\fill}

\begin{table}[h]
    \centering
    \begin{tabular}{||c c c c||}
 \hline
 Name & Fit point I & Fit point II & Data \\ [0.5ex]
 \hline
 $m_u$ ($M_Z$) [MeV] & 1.26673 & 1.24565 & 1.23 $\pm$ 0.21 \\
 $m_c$ ($M_Z$) [GeV] & 0.62797 & 0.62638 & 0.62 $\pm$ 0.017 \\
 $m_t$ ($M_Z$) [GeV] & 168.249 & 168.151 & 168.26 $\pm$ 0.75 \\
 $m_d$ ($M_Z$) [MeV] & 2.64273 & 2.67098 & 2.67 $\pm$ 0.19 \\
 $m_s$ ($M_Z$) [MeV] & 53.1172 & 53.3361 & 53.16 $\pm$ 4.61 \\
 $m_b$ ($M_Z$) [GeV] & 2.84242 & 2.84239 & 2.839 $\pm$ 0.026 \\
 $m_e$ ($M_Z$) [MeV] & 0.48316 & 0.48289 & 0.48307 $\pm$ 0.00045 \\
 $m_\mu$ ($M_Z$) [GeV] & 0.101765 & 0.101766 & 0.101766 $\pm$ 0.000023 \\
 $m_\tau$ ($M_Z$) [GeV] & 1.728544 & 1.728578 & 1.72856 $\pm$ 0.00028 \\
 \hline

 $|V_{ud}|$ & 0.973997 & 0.973993 & 0.9737 $\pm$ 0.00014 \\
 $|V_{us}|$ & 0.226529 & 0.226545 & 0.2245 $\pm$ 0.0008 \\
 $|V_{ub}|$ & 0.003769 & 0.003769 & 0.00382 $\pm$ 0.00024 \\
 $|V_{cd}|$ & 0.226416 & 0.226437 & 0.221 $\pm$ 0.004 \\
 $|V_{cs}|$ & 0.973174 & 0.973206 & 0.987 $\pm$ 0.011 \\
 $|V_{ct}|$ & 0.040843 & 0.039961 & 0.041 $\pm$ 0.0014 \\
 $|V_{td}|$ & 0.008076 & 0.007968 & 0.008 $\pm$ 0.0003 \\
 $|V_{ts}|$ & 0.040214 & 0.039341 & 0.0388 $\pm$ 0.0011 \\
 $|V_{tb}|$ & 0.999053 & 0.999126 & 1.013 $\pm$ 0.03 \\
 \hline
 $sin^2\theta_{12}$ & 0.31939 & 0.31358 & 0.31 $\pm$ 0.013 \\
 $sin^2\theta_{13}$ & 0.02246 & 0.02221 & 0.0224 $\pm$ 0.00066 \\
 $sin^2\theta_{23}$ & 0.58544 & 0.57928 & 0.582 $\pm$ 0.019 \\
 $\Delta m^2_{21} \times 10^{5}$ [eV] & 7.47418 & 7.44377 & 7.39 $\pm$ 0.21 \\
 $\Delta m^2_{31} \times 10^{3}$ [eV] & 2.55313 & 2.55596 & 2.525 $\pm$ 0.033 \\
 $\Delta m^2_{32} \times 10^{3}$ [eV] & 2.47839 & 2.48152 & 2.525 $\pm$ 0.033 \\
 \hline
 $M_W$ [GeV] & 80.391 & 80.392 & 80.433 $\pm$ 0.009 \cite{CDF:2022hxs} \\
 $M_{Higgs}$ [GeV] & 125.099 & 125.105 & 125.10 $\pm$ 0.14  \\
 \hline
 $\alpha_s$ & 0.1175 & 0.1173 & 0.1179 $\pm$ 0.001 \\
 $\alpha^{-1}_{em}$ & 127.950 & 127.951 & 127.952 $\pm$ 0.009 \\
 \hline
\end{tabular}
\captionsetup{justification=centering}
\caption{Low energy fitted observables at the $M_Z$ energy scale}
\label{Table:2}
\end{table}

\begin{table}
\centering
\begin{tabular}{||c|c|c|c||}
 \hline
 Name & Fit point I & Fit point II & Data \\ [0.5ex]
 \hline
 $\Delta a_\mu \times 10^{9}$  & 2.5659 & 2.5702 & 2.51 $\pm$ 0.59 \\
 $d_e \times 10^{31}$ & 4.0164 & 1.1198 & $< \pm 1.1 \times 10^2$ \\
 $d_\mu \times 10^{23}$ & -1.7168 & -1.6484 & $< \pm 1.8 \times 10^{4}$ \\
 \hline
 Re(C$^\mu_9$) & -0.02318 & -0.00694 & 0.0 $\pm$ 0.1 \\
 Im(C$^\mu_9$) & -0.00115 & -0.00929 & 0.0 $\pm$ 0.1 \\
 Re(C$^\mu_{10}$) $\times 10^{6}$ & 1.08887 & -2.85416 & 0.0 $\pm$ $10^5$ \\
 Im(C$^\mu_{10}$) $\times 10^{6}$ & 0.05418 & -3.81943 & 0.0 $\pm$ $10^5$ \\
 Re(C$^{\prime \mu}_9$) & 0.09117 & 0.07697 & 0.0 $\pm$ 0.1 \\
 Im(C$^{\prime \mu}_9$) & 0.03867 & 0.04170 & 0.0 $\pm$ 0.1 \\
 Re(C$^{\prime \mu}_{10}$) $\times 10^{5}$ & -0.42833 & 3.16388 & 0.0 $\pm$ $10^4$ \\
 Im(C$^{\prime \mu}_{10}$) $\times 10^{5}$ & -0.18162 & 1.71403 & 0.0 $\pm$ $10^4$ \\
 \hline
 BR($\mu \longrightarrow e \gamma$) $\times 10^{13}$ & 1.107 & 0.07124 & $<$ 4.2 \\
 BR($\tau \longrightarrow \mu \gamma$) $\times 10^{19}$ & $2.20265$ & $0.62282$ & $< 4.4 \times 10^{12}$ \\
 BR($\tau \longrightarrow \mu \mu \mu$) $\times 10^{20}$ & $9.87793$ & $2.76566$ & $< 2.1 \times 10^{13}$\\
 BR($K_L \longrightarrow \mu \mu$) $\times 10^{9}$ & 1.124679 & 0.080797 & $\leq$ 2.5 \\
 $R_{B_s \longrightarrow \mu \mu}$ & 0.87797 & 0.87777 & 0.75 $\pm$ 0.16  \\
 $\delta M_{12}$ & -0.016\% & -0.0016\% & $<\pm 6\%$ \\
 \hline
 $\sum_{i = 1}^3 m_{\nu_i}$ [eV] & 0.07531 & 0.07622 &  $< 0.12$ \\
 $\delta_{CP}$ (in radians) & -0.6713 & -0.7493 & $[-3.41, -0.03]$ \cite{T2K:2019bcf} \\
 \hline

\end{tabular}
\captionsetup{justification=centering}
 \caption{Low energy predicted observables}
 \label{Table:3}
\end{table}
\pagebreak

\begin{table}
    \centering
    \begin{tabular}{||c|c|c|c||}
    \hline
       Name  & Fit point I & Fit point II & Bound \\ [0.5ex]
       \hline
       $M_{Z'}$ (in TeV) &  1.478 & 1.481 & \\
       \hline
       $m_{\nu_4}$ (in TeV) & 1.881 & 1.867 & \\
       $m_{\nu_5}$ (in TeV) & 1.881 & 1.867 & \\
       \hline
       $m_{E_1}$ (in TeV) & 1.877 & 1.863 & $>$ 450 GeV \\
       $m_{E_2}$ (in TeV) & 1.881 & 1.867 & $>$ 450 GeV\\
       $m_{U_1}$ (in TeV) & 2.845 & 2.797 & $>$ 1.54 TeV \\
       $m_{U_2}$ (in TeV) & 3.026 & 3.028 & $>$ 1.54 TeV \\
       $m_{D_1}$ (in TeV) & 3.026 & 3.028 & $>$ 1.56 TeV \\
       $m_{D_2}$ (in TeV) & 3.650 & 3.626 & $>$ 1.56 TeV \\
       \hline
    \end{tabular}
    \captionsetup{justification=centering}
    \caption{New $Z'$ gauge boson, Dirac neutrinos, VL quark, and lepton masses}
    \label{Table:4}
\end{table}

We see that the two fit points fit the known experimental low energy data pretty nicely with  $\chi^2$ values of $23.02$ and $21.54$ if one considers the $W$ boson mass measurement of \cite{ParticleDataGroup:2018ovx} and $\chi^2$ values of $43.91$ and $41.32$ for the recent $W$ boson mass measurement of \cite{CDF:2022hxs}. It should be noted that the deviation of the $W$ boson mass from its SM value predicted in our model is owing to the fact that we have considered the experimental data of \cite{CDF:2022hxs} to fit, and if one instead considers the $W$ boson mass measurements of \cite{ParticleDataGroup:2018ovx,ATLAS:2023fsi}, then these can be accommodated in the model. Moreover, with the consideration of the $W$ boson mass of \cite{CDF:2022hxs}, the value predicted by our fit points is not within 1$\sigma$ of this experimental measurement, and hence our fit points predict a value in between the SM prediction and the measurement \cite{CDF:2022hxs}. Thus we are not able to fit the new $W$ mass measurement of \cite{CDF:2022hxs} within 1$\sigma$. They also solve the muon $g$ - 2 with all Wilson coefficients consistent with recent experimental data. Moreover, as can be seen in \textbf{Table 4} below, the new VL quarks and leptons predicted by the model all have masses at the TeV scale consistent with the experimental bounds in \cite{68} and \cite{CMS:2022cik}. The new $Z'$ gauge boson and the two Dirac neutrinos also have a mass of $\approx 1$ TeV and hence all these particles are within reach of future experiments. Although we do not consider a detailed analysis for $M_{Z^\prime}$ here, the mass of the $Z^\prime$ boson in our model $M_{Z^\prime} \sim $1.5 TeV, is in the range of current LHC limits, and is consistent with the $M_{Z^\prime} \gtrsim 1.3$ TeV bound corresponding to $g^\prime \sim 0.27$ in \cite{Alonso:2017uky} which is realized in our model. Moreover, due to $(\hat{g}^d_{L})_{23}$ being very small i.e. $\mathcal{O}(10^{-4})$, an $M_{Z^\prime}$ of 1.5 TeV is also not excluded due to constraints obtained in more simplistic models \cite{Allanach:2019mfl, Bonilla:2017lsq} and where lighter $Z^\prime$ bosons have been considered \cite{Kohda:2018xbc}. Hence, this $Z^\prime$ boson can be searched for at the LHC. The model also predicts a value of $sin(\delta_{CP}) = -0.622$ and the Jarlskog invariant $J = -0.0276$ for the fit point I and $sin(\delta_{CP}) = -0.681$ and the Jarlskog invariant $J = -0.0277$ for the fit point II, consistent with the $3\sigma$ confidence interval of $\delta_{CP} \in [-3.41, -0.03]$ found in \cite{T2K:2019bcf}.

\subsection{SUSY mass spectrum}
Our model predicts all the SUSY particles to have masses less than 25 TeV. The sparticle mass parameters for all quarks and leptons have been presented in the \hyperref[sec:7.5.3]{Appendix~\ref*{sec:7.5.3}} and \hyperref[sec:7.8.3]{Appendix~\ref*{sec:7.8.3}}. We consider the Generalized Mirage Mediation (GMM) model and choose a universal $c_m$ for all the sparticle mass parameters except the $c_{H_u}$ and $c_{H_d}$ Higgs parameters. With this choice, the RG flow of these mass parameters can be analyzed for the squarks and sleptons and it can be seen how they vary as the energy scale is reduced. Working within the GMM model, our model necessarily requires that the mass scale of the SUSY particles is not significantly smaller than $\sim$ 3 TeV, because the gluino mass bound of $m_{gluino} \gtrsim 1.9$ TeV would be violated if the parameters of the GMM model are significantly smaller than those in \hyperref[sec:7.5.3]{Appendix~\ref*{sec:7.5.3}} and \hyperref[sec:7.8.3]{Appendix~\ref*{sec:7.8.3}}. On the other hand, in principle, one can have masses of the SUSY particles larger than $\sim$ 30 TeV of our model, which would, however further suppress the SUSY contributions to branching ratios discussed above and also increase $\mu$ further.

\textbf{Figure 2} and \textbf{Figure 3} show the RG flows (for fit points I and II respectively) of the sparticle mass parameters $m_{squarks}$ and $m_{sleptons}$ from the GUT scale (which is $\approx 5 \times 10^{16}$ GeV) to the weak scale $M_Z$. We have plotted the sparticle mass parameters for the 3rd, $4$, and $\bar 4$ families. Despite the messy appearance, they tell us that on average, the squark mass parameters grow much faster as compared to the slepton mass parameters, and hence, on average, the values of the squark mass parameters are more than that of the slepton mass parameters at the weak scale $M_Z$. This is similar to what happens in the MSSM. We can see that the soft SUSY breaking mass parameters are all less than 24 TeV and, since the VL quark and lepton masses are less than 5 TeV, the scalar sparticle masses of all the squarks and sleptons are less than 25 TeV.

We have also plotted the RG flow of the gaugino mass parameters $M_a$ in \textbf{Figure 4} and \textbf{Figure 5} for fit points I and II respectively. The plots show an important difference between the MSSM and our supersymmetric model: Along with the gaugino mass parameters for the U(1) and SU(2) gauge groups, we see that the gluino mass parameter also decreases as we RG flow to the weak scale. This is however expected because unlike the MSSM, the strong SU(3) coupling increases with energy as can be seen in \textbf{Figure 6} in the Appendix.\footnote{Note, we have demanded precise gauge coupling unification in our analysis.}  The new $Z'$ gaugino corresponding to the additional U(1)$^\prime$ is the LSP of the model. From the plots, we see that the gaugino mass parameters (especially $M_1$, $M_2$ and $M^\prime$) sharply fall from $>$ 4 TeV to less than 1 TeV as we RG flow from the GUT scale to the weak scale. The fall is the sharpest for the gaugino mass parameter $M^\prime$. The model also predicts a gluino mass $m_{gluino} = 2.3749$ TeV for the fit point I and $m_{gluino} = 3.0332$ TeV for the fit point II, both of which are consistent with the constraint $m_{gluino} \gtrsim 1.9 $ TeV.

The chargino and neutralino masses, calculated by diagonalizing the neutralino mass matrices given in the \hyperref[sec:7.3]{Appendix~\ref*{sec:7.3}}, are summarized in \textbf{Table 5}. We see that the two fit points give a light neutralino mass ranging from $\approx$ 480 - 580 GeV. From the SUSY mass spectrum, i.e., the sparticle mass parameters and the chargino, neutralino masses, we get that this neutralino is indeed the lightest supersymmetric particle (LSP). Within the approximations of \cite{Wells:1997ag}, we have checked that the SUSY mass scale ($\sim 10$ TeV) and the LSP mass predicted by our model yields an $\Omega_{\chi} h^2$ value of less than 0.12, consistent with relic density constraints and preventing overclosing of the universe. This makes it possible for this neutralino to be the dark matter candidate.
\begin{figure}[]
    \centering
    \includegraphics[width=10cm]{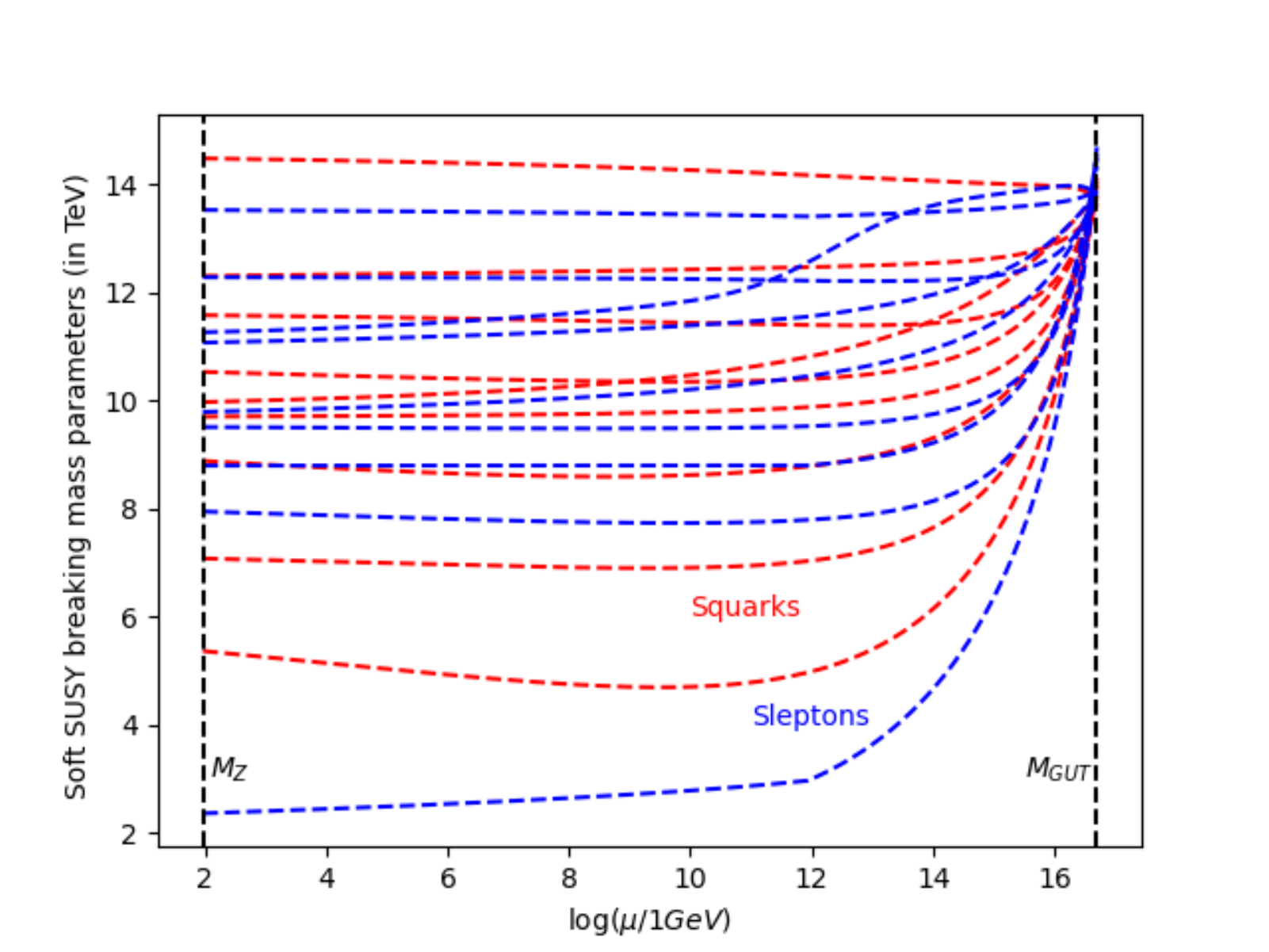}
    \caption{RG flow of the soft SUSY breaking mass parameters for squarks and sleptons (fit point I)}
    \label{fig:my_label}
\end{figure}

\begin{figure}[]
    \centering
    \includegraphics[width=10cm]{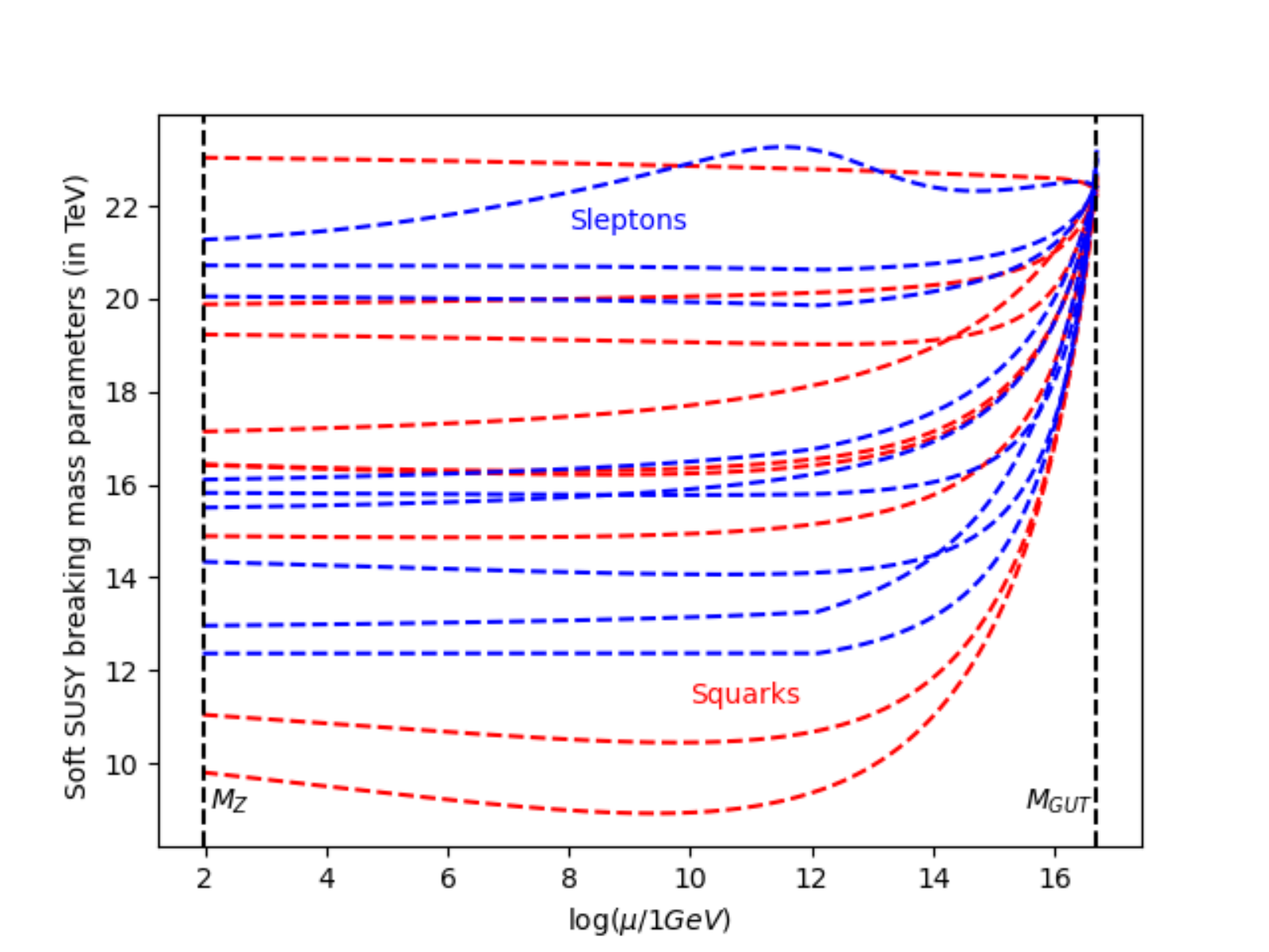}
    \caption{RG flow of the soft SUSY breaking mass parameters for squarks and sleptons (fit point II)}
    \label{fig:my_label}
\end{figure}

\begin{figure}[]
    \centering
    \includegraphics[width=10cm]{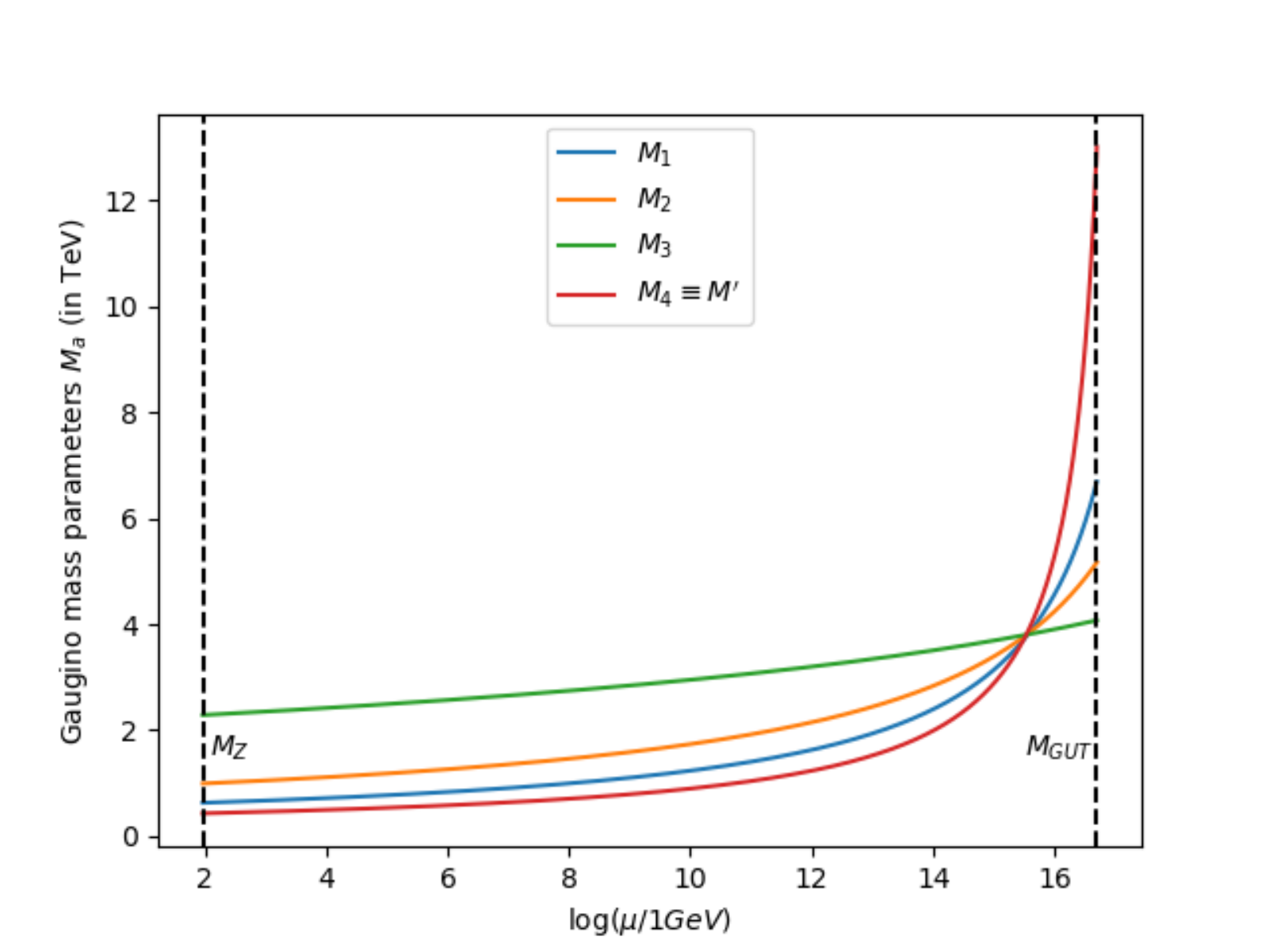}
    \captionsetup{justification=centering}
    \caption{RG flow of the gaugino mass parameter $M_a$ (fit point I)}
    \label{fig:my_label}
\end{figure}

\begin{figure}[]
    \centering
    \includegraphics[width=10cm]{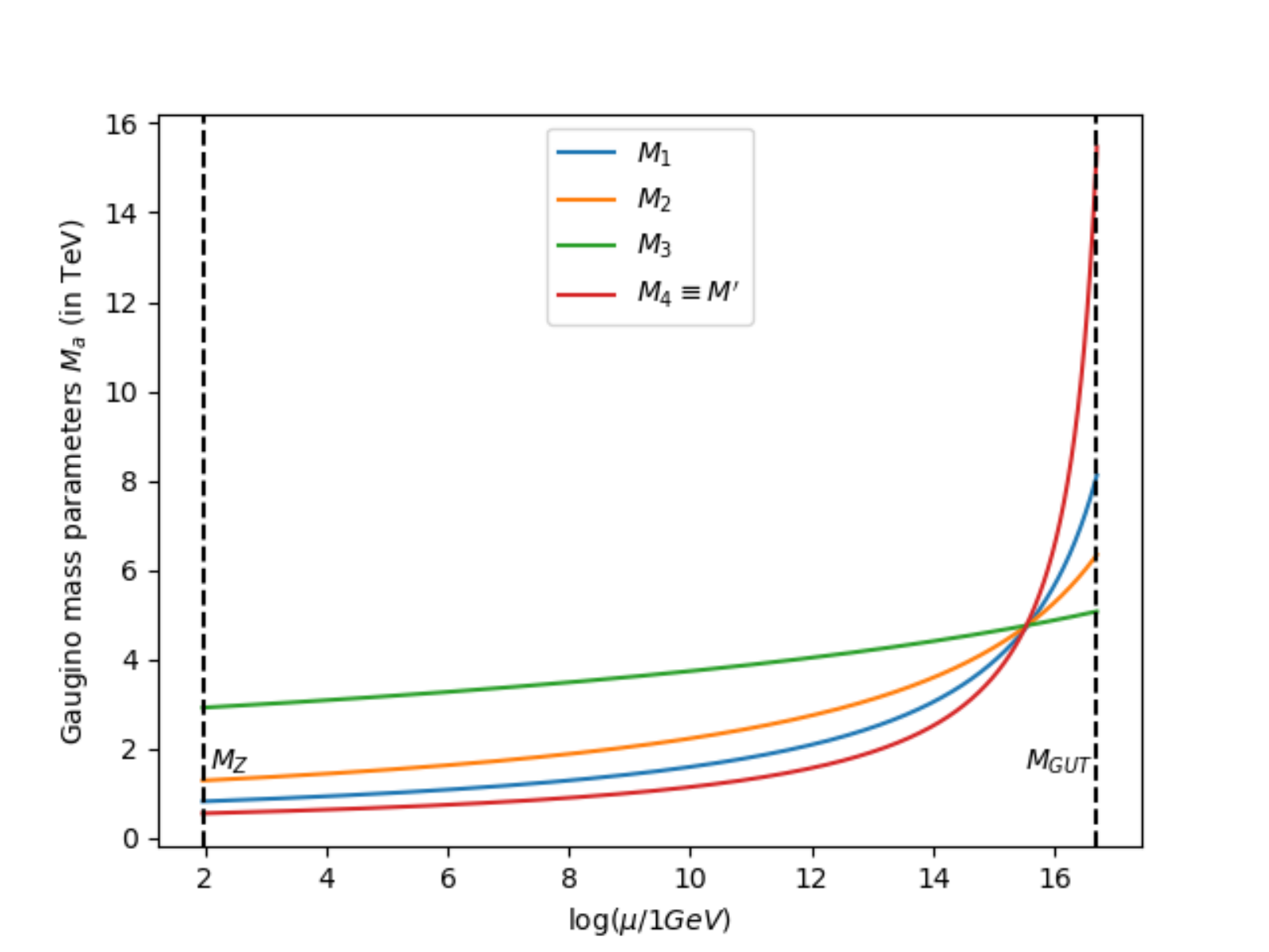}
    \captionsetup{justification=centering}
    \caption{RG flow of the gaugino mass parameter $M_a$ (fit point II)}
    \label{fig:my_label}
\end{figure}

\begin{table}[]
    \centering
    \begin{tabular}{||c|c|c||}
    \hline
     Name & Fit point I & Fit point II  \\
     \hline
    $m_{\Tilde{C}_1}$ (in TeV) & 0.992 & 1.2798  \\
    $m_{\Tilde{C}_2}$ (in TeV) & 13.558 & 22.147 \\
    \hline
    $m_{\Tilde{N}_1}$ (in GeV) & 480.41 & 578.52  \\
    $m_{\Tilde{N}_2}$ (in GeV) & 628.08 & 811.98  \\
    $m_{\Tilde{N}_3}$ (in TeV) & 0.992 & 1.2798  \\
    $m_{\Tilde{N}_4}$ (in TeV) & 13.558 & 22.147  \\
    $m_{\Tilde{N}_5}$ (in TeV) & 13.558 & 22.147  \\
    $m_{\Tilde{N}_6}$ (in TeV) & 19.824 & 29.902  \\
    $m_{\Tilde{N}_7}$ (in TeV) & 19.875 & 29.936  \\
    \hline
    \end{tabular}
    \captionsetup{justification=centering}
    \caption{Chargino and neutralino masses at $M_Z$ energy scale}
    \label{tab:my_label}
\end{table}

\section{CONCLUSION}
In this paper, we have studied the Standard Model extended by including supersymmetry and a complete family of vector-like fermions, including right-handed neutrinos which are vector-like with respect to the Standard Model gauge group. In addition to that, we have introduced a new $U(1)_{3-4} \equiv U(1)^\prime$ gauge group under which the third SM family of quarks and leptons has a U(1)$_{3-4}$ charge of 1 while the left chiral part of the 4th family has a charge of -1. This model is also free of gauge anomalies. In this paper, we have considered a complete supersymmetric analysis of the 4$^{th}$ vector-like chiral family along with the 3 SM families.

Our supersymmetric model unifies the four gauge couplings at $M_{GUT} \approx 5 \times 10^{16}$ GeV. We have also shown that our model can explain the anomalies in $g$ - 2 without conflicting with other low-energy experimental data. The model also predicts the new VL quarks, leptons, 2 Dirac neutrinos and the $Z^{'}$ gauge boson to have a mass at the TeV scale and also predicts the branching ratios of $\mu \longrightarrow e \gamma$, $\tau \longrightarrow \mu \gamma$ and $\tau \longrightarrow \mu \mu \mu$ with $\mu \longrightarrow e \gamma$ within the reach of future experiments, but with BR($\tau \longrightarrow \mu \gamma$) and BR($\tau \longrightarrow \mu \mu \mu$) highly suppressed.
\\

The model predicts a mass spectrum of the supersymmetric particles with all sparticle masses $<$ 25 TeV and also gives a light neutralino $\approx 480$ GeV for the fit point I and $\approx 578$ GeV for the fit points II, which is a potential dark matter candidate. Hence, we can see that indeed the "vector-like chiral" model of \cite{Raby:2017igl} can be made into a supersymmetric grand-unified model which not only solves the muon anomalies but also explains the deviation of other low energy data from the values predicted by the Standard Model. Moreover, the model predictions and the explained anomalies are not sensitive to the neutrino sector parameters and reasonable changes to these parameters do not change the model predictions and the low energy observables of the other sectors significantly. Hence, we do not rule out the possibility that our model predictions and results might also be obtained by a different fit point with reasonably different neutrino sector Yukawa and mass parameters.

Due to the vector-like family which has been introduced in our model, the GUT gauge coupling constant is larger than in a theory without a VL family.  In addition the GUT scale is also larger.   This will have consequences for dimension six proton decay into the dominant decay mode, $p \rightarrow e^+ \pi^0$.  However using the results from the \hyperref[sec:7.5]{Appendix~\ref*{sec:7.5}} we find that the lifetime in this model is only a factor of about 0.4 shorter than in a $SU(5)$ SUSY GUT without a VL family.   A more detailed analysis of proton decay is beyond the scope of the present paper.  For a detailed calculation of proton decay in a model with additional VL $SU(5)$ $5$ and $\bar 5$, see Ref. \cite{Hisano:2017spq}.

\section{APPENDIX}
\label{sec:Appendix}
\subsection{Gauge coupling unification}
Our model unifies the 4 gauge couplings of the $SU(3)_C \times SU(2)_L \times U(1)_Y \times U(1)_{3-4}$ at a GUT scale of $M_G \approx 5 \times 10^{16}$ GeV. The plot for the two-loop RG flow of the 4 gauge couplings is shown in the \textbf{Figure 6}.
\begin{figure}
    \centering
    \includegraphics[width=10cm]{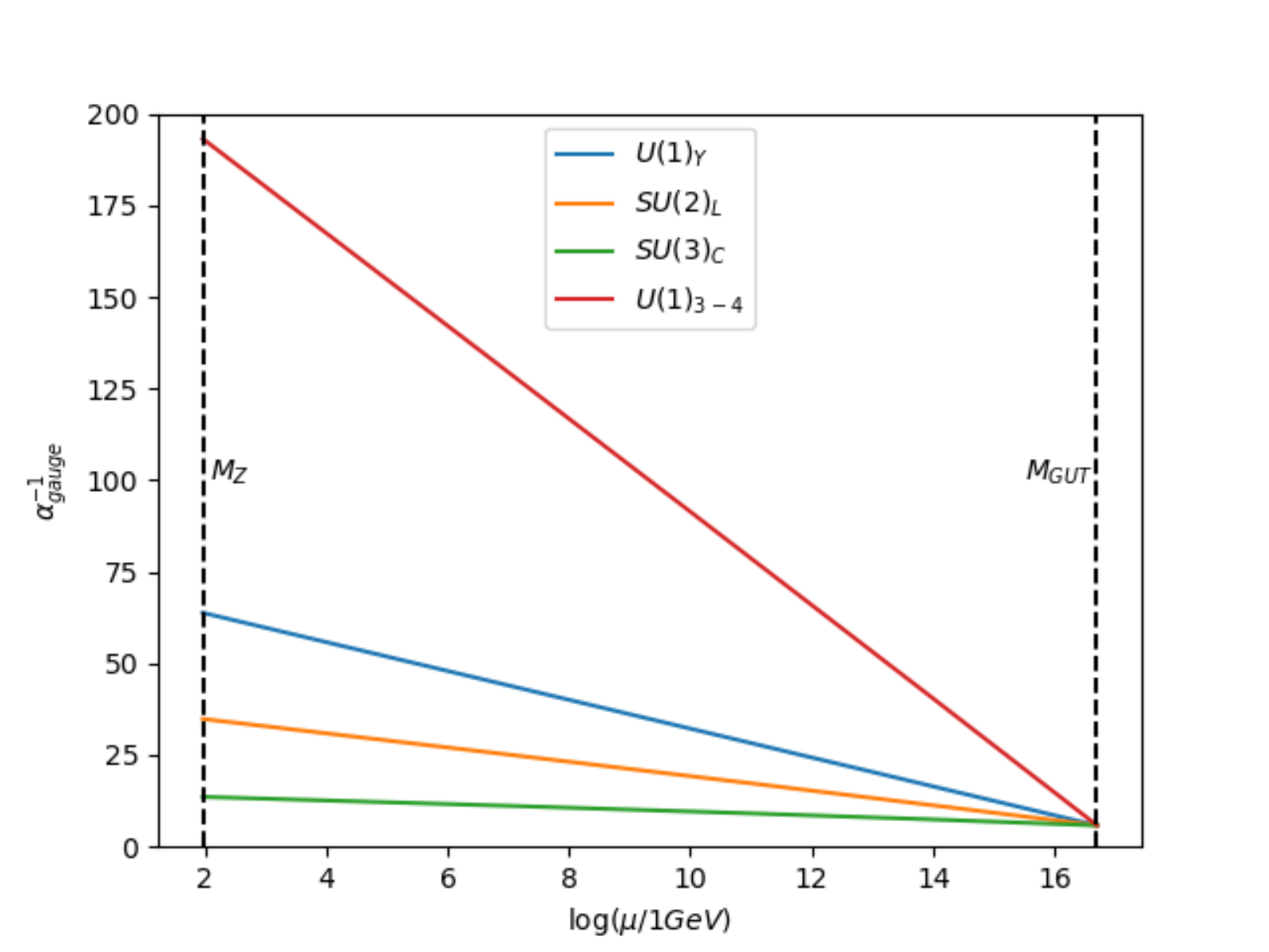}
    \captionsetup{justification=centering}
    \caption{RG flow of gauge couplings inverse ($\alpha^{-1}_{gauge}$) as a function of energy scale}
    \label{fig:my_label}
\end{figure}
At the GUT scale $M_G$, we have:
\begin{equation}
    \Tilde{\alpha}_G = \alpha_1(M_G) = \alpha_2(M_G), \ \epsilon_3 = \frac{\alpha_3(M_G) - \Tilde{\alpha}_G}{\Tilde{\alpha}_G}
\end{equation}
where the parameter $\epsilon_3$ in our model is zero, consistent with the expectation when SU(5) gauge group is broken down to SM gauge group via a Wilson line in an extra dimension. The model achieves unification of all the 4 gauge couplings at $M_G$ and additionally solves the low energy anomalies. Since the RG flow analysis is at the two-loop level, we also consider one-loop threshold corrections to get the gauge couplings actually observed at low energies:
\begin{equation}
    \frac{1}{\alpha_i(M_Z)_{\overline{MS}}} = \frac{1 - \alpha_i(M_Z)_{\overline{DR}}\Delta \alpha_i}{\alpha_i(M_Z)_{\overline{DR}}} + \frac{C_2(G_i)}{12\pi}
\end{equation}
where $C_2(G_i)$ = N for SU(N) and = 0 for U(1). $\Delta \alpha_i$ is the threshold correction to the $\alpha_i(M_Z)_{\overline{DR}}$ that depends on the sparticle and VL particle masses at the weak scale primarily. With the 2 loop RG equations and 1 loop threshold corrections taken into account, it gives the value of $g'$ at the weak scale to be $0.2673$ for fit point I and 0.2680 for fit point II. This is consistent with the bound of $g^\prime < 0.271$ found from the 1 loop RG equations.

It is also worth noting that, in contrary to the SM and the MSSM, the $g_3$ coupling of $SU(3)_C$ in this model is \textit{not} asymptotically free and instead increases with energy, much like the other three gauge couplings.

\subsection{Integrating out the heavy neutrino states}
\label{sec:7.1}
From the Lagrangian of our model, we can integrate out the heavy states. The equations of motion for the heavy fields give us:
\begin{align*}
 -{\delta \mathcal{L} \over \delta \overline{ N_L^C}} & = M_L N_L  + \lambda_{RLN} {\bar L}_L H_d + \lambda_{2N_a} {\bar \nu^{a}}_L + \lambda_{3 N} \Phi {\bar \nu^{3}}_L + \lambda_N \bar \Phi {\bar N}_L & = 0 \\
 -{\delta \mathcal{L} \over \delta \overline{ {\bar N}_L^C}} & =  M_R {\bar \nu}^3_L   + \lambda_{LRN} L_L H_u + \bar \Phi \lambda_N N_L                     & = 0 \\
 -{\delta \mathcal{L} \over \delta \overline{{\bar \nu}_L^{3}{}^C}} & =  M_R {\bar N}_L  + y^\nu_{33} l^{3}_L H_u + \lambda_{3 N} \Phi N_L       & = 0  \\
 -{\delta \mathcal{L} \over \delta \overline{{\bar \nu}_L^{a}{}^C}} & =  M_R^{a b} {\bar \nu}^b_L  + y^\nu_{a b} l^{b}_L H_u + \lambda_{2 N a} N_L       & = 0
 \end{align*}
 Then, solving for the heavy fields, we get:
 \begin{align*}
  N_L & =  - \frac{1}{M_L} \left( \lambda_{RLN} {\bar L}_L H_d + \lambda_{2N_a} {\bar \nu}^{a}_L + \lambda_{3 N} \Phi {\bar \nu^{3}}_L + \lambda_N \bar \Phi {\bar N}_L\right) & \\
  &  \sim - \frac{1}{M_L} \left( \lambda_{RLN} {\bar L}_L H_d \right) & \\
{\bar \nu}^3_L & = - \frac{1}{M_R} \left( \lambda_{LRN} L_L H_u + \bar \Phi \lambda_N N_L \right)  & \\
 & \sim   - \frac{1}{M_R} \left( \lambda_{LRN} L_L H_u \right) &  \\
 {\bar N}_L & = -\frac{1}{M_R} \left( y^\nu_{33} l^{3}_L H_u + \lambda_{3 N} \Phi N_L\right)  & \\
  & \sim -\frac{1}{M_R} \left( y^\nu_{33} l^{3}_L H_u \right)  &   \\
{\bar \nu}^d_L & =  - {M_R^{-1}}^{d a}  \left( y^\nu_{a b} l^{b}_L H_u + \lambda_{2 N a} N_L\right) & \\
 & \sim - {M_R^{-1}}^{d a}  \left( y^\nu_{a b} l^{b}_L H_u \right) &
 \end{align*}
 where we have neglected terms suppressed by two powers of the heavy mass scale. \\
 Now plugging these back into the original Lagrangian, we obtain the effective Lagrangian for just the light neutrino states:
 \begin{align*}
    \mathcal{L}^{eff} &=  \overline{(l^{d}_L)^C} (y^\nu_{d e})^T H_u (M_R^{-1})_{e a} y^\nu_{db} l^b_{L}H_u  +  \frac{1}{M_R} \lambda_{LRN}  (y^\nu_{33}) \overline{(l^{3}_L)^C} H_u  L_L H_u & \\ & - \overline{{\bar L}_L^C} \left(  \lambda_{2L_a} l_L^a + \lambda_{3 L} \bar \Phi l^3_L + \lambda_L \Phi L_L   \right) +   \frac{1}{M_L} \lambda_{RLN} \lambda_{RLN} \overline{{\bar L}_L^C} H_d  \bar{L}_L H_d  + h.c. &
\end{align*}
This effective Lagrangian is considered for RG flow and for the physics below the right-handed Majorana neutrino mass scale.

\subsection{Radiative corrections to the Higgs mass}
\label{sec:7.2}
In our model, we consider the large $tan\beta$ regime with $\tan\beta \sim 47.6$. In this regime, we get $|\mu|^2$ at the tree level as:
\begin{equation}
    |\mu|^2 = \frac{1}{2}(-2m^2_{H_u} - m_Z^2 + \frac{2}{tan^2\beta}(m^2_{H_d} - m^2_{H_u}))
\end{equation}
However, in our analysis, we consider radiative corrections to the $Z$ boson mass, which reduces the $\mu$ further to approximately give:
\begin{equation}
    |\mu|^2 \approx -m^2_{H_u} - \frac{m_Z^2}{2} - \Sigma_u^u
\end{equation}
where $\Sigma^u_u$ corresponds to the radiative corrections and its exact expression is given in the Appendix of \cite{Baer:2012cf}. We also consider the contribution of the VL squarks along with those mentioned in \cite{Baer:2012cf}.
Including the large radiative corrections from top quark, top squark \cite{Martin:1993zk} and VL particle masses \cite{Endo:2011mc}, the Higgs mass is given by:
\begin{equation}
    m^2_{h^0} = \frac{1}{2}(m^2_{A^0} + m^2_Z - \sqrt{(m^2_{A^0}-m_Z^2)^2 + 4m^2_Zm^2_{A^0}sin^2(2\beta)}) + \Delta_{t}(m^2_{h^0}) + \Delta_{VL}(m^2_{h^0})
\end{equation}
where
\begin{equation}
    \Delta_t(m^2_{h^0}) = \frac{3}{4\pi^2}cos^2\alpha |y_t|^2 m_t^2 [ln(m_{\Tilde{t}_1}m_{\Tilde{t}_2}/m_t^2)] + \Delta_{threshold}
\end{equation}
with
\begin{equation}
    \Delta_{threshold} = c_{\Tilde{t}}^2s_{\Tilde{t}}^2[(m^2_{\Tilde{t}_2} - m^2_{\Tilde{t}_1})/m_t^2]ln(m^2_{\Tilde{t}_2}/m^2_{\Tilde{t}_1}) + c_{\Tilde{t}}^4s_{\Tilde{t}}^4[(m^2_{\Tilde{t}_2} - m^2_{\Tilde{t}_1})^2 - \frac{1}{2}(m_{\Tilde{t}_2}^4 - m_{\Tilde{t}_1}^4)ln(m^2_{\Tilde{t}_2}/m^2_{\Tilde{t}_1})]/m_t^4
\end{equation}
and
\begin{equation}
    \Delta_{VL}(m^2_{h^0}) \approx -\frac{3v^2}{48 \pi^2}|\lambda_{RLU}|^4 \frac{|\mu|^4}{m_{S_U}^4} - \frac{3v^2}{48 \pi^2}|\lambda_{RLD}|^4 \frac{|\mu|^4}{m_{S_D}^4}
\end{equation}
because in our model we have $|\lambda_{RLU}| > |\lambda_{LRU}|$ and $|\lambda_{RLD}| > |\lambda_{LRD}|$. \\
The angle $\alpha$ is determined by:
\begin{equation}
    \frac{tan(2\alpha)}{tan(2\beta)} = \frac{m^2_{A^0} + m_Z^2}{m^2_{A^0} - m_Z^2}
\end{equation}
$m_{\Tilde{t}_1}$ and $m_{\Tilde{t}_2}$ are the top squark masses, $m_{S_U}$ and $m_{S_D}$ is the scale of VL scalar stop and sbottom masses and $m^2_{A^0} = 2|\mu|^2 + m^2_{H_u} + m^2_{H_d}$. \\
\\
With $tan\beta \sim 47.6$, this model gives a reasonable $\mu \approx 13.6$ TeV for the fit point I and  $\mu \approx 22.1$ TeV for the fit point II. The Higgs mass can be easily fine tuned by varying $c_{H_u}$ and $c_{H_d}$ appropriately.

\subsection{The potential for the new scalar fields and self-consistency equations}
The classical scalar potential for the $\Phi$ and $\bar \Phi$ fields is given by:
\begin{equation}
    V(\Phi, \bar \Phi) = (|\mu_{\Phi}|^2 + m^2_{\Phi})|\Phi|^2 + (|\mu_{\Phi}|^2 + m^2_{\bar \Phi})|\bar \Phi|^2 - (b_{\Phi} \mu_{\Phi} \Phi \bar \Phi + c.c) + \frac{1}{2} {g^{\prime}}^2(|\Phi|^2 - |\bar \Phi|^2)^2
\end{equation}
It can be easily seen that this potential is indeed bounded from below. To see this, note that, except for the D-flat directions i.e. when $|\Phi| \neq |\bar \Phi|$, the above potential is clearly bounded from below since the term quartic in fields dominates for large values of $|\Phi|$ and $|\bar \Phi|$. Now if $|\Phi| = |\bar \Phi|$, then the condition for $V(\Phi, \bar \Phi)$ to be bounded from below is given by:
\begin{equation}
    2|\mu_\Phi|^2 + m_\Phi^2 + m_{\bar \Phi}^2 > 2 b_\Phi \mu_\Phi
\end{equation}
Minimizing the scalar potential:
\begin{equation}
    \frac{\partial V}{\partial \Phi} = \frac{\partial V}{\partial \bar \Phi} = 0
\end{equation}
at $v_\Phi$ and $v_{\bar \Phi}$ and considering real VEVs $v_\Phi, v_{\bar \Phi}$, we get the two self-consistency equations:
\begin{equation}
    (|\mu_\Phi|^2 + m^2_\Phi)v_\Phi - b_\Phi \mu_\Phi v_{\bar \Phi} - {g^\prime}^2 v_\Phi(v_{\bar \Phi}^2 - v_\Phi^2) = 0
\end{equation}
and
\begin{equation}
    (|\mu_\Phi|^2 + m^2_{\bar \Phi})v_{\bar \Phi} - b_\Phi \mu_\Phi v_{\Phi} + {g^\prime}^2 v_{\bar \Phi}(v_{\bar \Phi}^2 - v_\Phi^2) = 0
\end{equation}
With $\mu_\Phi = 19.8232$ TeV and $b_\Phi = 11.6600$ TeV for the fit point I and $\mu_\Phi = 29.9013$ TeV and $b_\Phi = 14.6578$ TeV for the fit point II, the above two self-consistency equations are satisfied with VEVs $v_\Phi$ and $v_{\bar \Phi}$ mentioned in the \hyperref[sec:7.5]{Appendix~\ref*{sec:7.5}} and \hyperref[sec:7.8]{Appendix~\ref*{sec:7.8}} for the two fit points respectively. Hence, the new U(1)$^\prime$ gauge symmetry is consistently broken by the VEVs of the new scalar fields.

\subsection{Chargino and Neutralino masses}
\label{sec:7.3}
In the basis ($\Tilde{Z}^0$, $\Tilde{\Phi}^0$, $\Tilde{\bar \Phi}^0$, $\Tilde{B}^0$, $\Tilde{W}^0$, $H^0_d$, $H^0_u$) (where $\Tilde{Z}^0$ is the gaugino corresponding to the new U(1)$^\prime$), the neutralino mass matrix is given by:
\begin{equation}
    \textbf{M}_{\Tilde{N}} = \begin{pmatrix}
    M' & -g'v_\Phi & -g'v_{\bar \Phi} & 0 & 0 & 0 & 0 & \\
    -g'v_\Phi & 0 & -\mu_\Phi & 0 & 0 & 0 & 0 & \\
    -g'v_{\bar \Phi} & -\mu_\Phi & 0 & 0 & 0 & 0 & 0 & \\
    0 & 0 & 0 & M_1 & 0 & -\sqrt{3/5}g_1v_{H_d}/\sqrt{2} & \sqrt{3/5}g_1v_{H_u}/\sqrt{2} & \\
    0 & 0 & 0 & 0 & M_2 & g_2v_{H_d}/\sqrt{2} & -g_2v_{H_u}/\sqrt{2} & \\
    0 & 0 & 0 & -\sqrt{3/5}g_1v_{H_d}/\sqrt{2} & g_2v_{H_d}/\sqrt{2} & 0 & -\mu & \\
    0 & 0 & 0 & \sqrt{3/5}g_1v_{H_u}/\sqrt{2} & -g_2v_{H_u}/\sqrt{2} & -\mu & 0

    \end{pmatrix}
\end{equation}
where $M'$, $M_1$, and $M_2$ are the weak scale values for the soft SUSY breaking gaugino mass parameters.
The mass eigenstates can be obtained by diagonalizing this mass matrix using a unitary matrix \textbf{N}:
\begin{equation}
    \Tilde{N}_i = \textbf{N}_{ij} \psi^0_j
\end{equation}
where $\Tilde{N}_i$ are the mass eigenstates. Hence we get the diagonal neutralino mass matrix as:
\begin{equation}
    \textbf{N} \textbf{M}_{\Tilde{N}} \textbf{N}^{-1} = \begin{pmatrix}
    m_{\Tilde{N}_1} & 0 & 0 & 0 & 0 & 0 & 0 & \\
    0 & m_{\Tilde{N}_2} & 0 & 0 & 0 & 0 & 0 &\\
    0 & 0 & m_{\Tilde{N}_3} & 0 & 0 & 0 & 0 &\\
    0 & 0 & 0 & m_{\Tilde{N}_4} & 0 & 0 & 0 &\\
    0 & 0 & 0 & 0 & m_{\Tilde{N}_5} & 0 & 0 &\\
    0 & 0 & 0 & 0 & 0 & m_{\Tilde{N}_6} & 0 &\\
    0 & 0 & 0 & 0 & 0 & 0 & m_{\Tilde{N}_7} &
    \end{pmatrix}
\end{equation}
Similarly, the chargino masses are given in terms of the $M_2$ gaugino mass and the Higgsino mixing parameter $\mu$ by:
\begin{equation}
    m^2_{\Tilde{C}_1}, m^2_{\Tilde{C}_2} = \frac{1}{2}[M_2^2 + |\mu^2| + 2m_W^2 \mp \sqrt{(M_2^2 + |\mu^2| + 2m_W^2)^2 - 4|\mu M_2 - m_W^2 sin(2\beta)|^2}]
\end{equation}
The predictions for these masses in our model are consistent with all the constraints and also imply that the lightest supersymmetric particle (LSP) of the model is the lightest neutralino. In our model, it has the potential to be a good candidate for dark matter.

\subsection{Input parameters at the fit point - Fit point I}
\label{sec:7.5}
\begin{equation}
    g_1 = g_2 = g_3 = g_4 = g_{GUT} = 1.4729, \ \epsilon_3 = 0, \ M_{GUT} = 5.012 \times 10^{16} \ GeV
\end{equation}
\begin{equation}
    tan\beta = 47.59696, \ v_\Phi = 3041.536 \ GeV, \ v_{\bar \Phi} = 2391.531 \ GeV
\end{equation}
\begin{equation}
    \lambda_{2Q_1} = 843.086 \ GeV, \ \lambda_{2Q_2} = 1149.109 \ GeV, \ \lambda_{2L_1} = \lambda_{2L_2} = 1300.727 \ GeV
\end{equation}
\begin{equation}
    \lambda_{2U_1} = 646.234 \ GeV, \ \lambda_{2U_2} = 881.402 \ GeV, \ \lambda_{2D_1} = 989.323 \ GeV, \lambda_{2D_2} = 3104.476 \ GeV
\end{equation}
\begin{equation}
    \lambda_{2E_1} = -1837.064 \ GeV, \ \lambda_{2E_2} = 0 \ GeV
\end{equation}
The remaining 50 input parameters are presented in the form of the Yukawa matrices at the GUT scale and the soft SUSY breaking parameters. Many of the entries in the Yukawa matrices are 0 and some of them are equal (in the GUT sense) which together give 38 of the model parameters. The Yukawa matrices at the GUT scale are:
\begin{equation} Y_{u} = \begin{pmatrix}  0.00000937 &  0.000000 & 0.000000 &  0.000000 &  0.000000 & \\ 0.000000 &  0.000000 &  0.000000 &  0.000000  &  0.000000 & \\ 0.000000 &  0.000000 &  1.898317 e^{ i0.789853} &  0.000000 &  1.705787 e^{ -i0.000153} &  \\ 0.000000 &  0.000000 &  0.000000 &  0.017169 & 1.616310 &  \\ 0.000000 &  0.000000 &  0.146814 e^{ -i0.038278} &  2.193911 &  2.163913 \\ \end{pmatrix} \end{equation}

\begin{equation} Y_{d} = \begin{pmatrix}  0.002213 &  0.006113 e^{ i1.193432} & 0.000000 &  0.000000 &  0.000000 & \\ 0.017042 e^{ i0.933044} &  0.043694 e^{ i0.768385} &  0.000000 &  0.000000  &  0.000000 & \\ 0.000000 &  0.000000 &  1.898317 e^{ i0.789853} &  0.000000 &  2.565781 e^{ -i0.000123} &  \\ 0.000000 &  0.000000  &  0.000000 &  0.000917 & 0.040027 &  \\ 0.000000 &  0.000000 &  0.146814 e^{ -i0.038278} &  2.193911 &  0.797559 \\ \end{pmatrix} \end{equation}

\begin{equation} Y_{e} = \begin{pmatrix}  0.000224  &  0.000188  & 0.000000  &  0.000000  &  0.000000  & \\ 0.000186  &  0.000000  &  0.000000  &  0.000000   &  0.000000  & \\ 0.000000  &  0.000000  &  1.898317 e^{ i0.789853} &  0.000000  &  0.000000  &  \\ 0.000000  &  0.000000  &  0.000000  &  0.081866  & 0.467347  &  \\ 0.000000  &  0.000000  &  0.000000  &  0.957657  &  2.959867  \\ \end{pmatrix} \end{equation}

\begin{equation} Y_{\nu} = \begin{pmatrix}  2.192903  &  2.315101  & 0.000000  &  0.000000  &  0.000000  & \\ 1.847680  &  1.891011  &  0.000000  &  0.000000   &  0.000000  & \\ 0.000000  &  0.000000  &  0.759619  &  0.000000  &  0.000000  &  \\ 0.000000  &  0.000000  &  0.000000  &  2.476346  & 1.943457  &  \\ 0.000000  &  0.000000  &  0.000000  &  0.957657  &  2.066375  \\ \end{pmatrix} \end{equation}

\subsubsection{Quark and charged lepton Yukawa matrices at the weak scale}

\begin{equation} Y_{\nu} = \begin{pmatrix}  0.00001628 e^{ i0.000001} &  0.000007 e^{ i0.000066} & 0.000000  &  0.000000  &  0.000000  & \\ 0.000000  &  0.000000  &  0.000000  &  0.000000   &  0.000000  & \\ 0.000000  &  0.000000  &  1.000457 e^{ i0.789853} &  0.000000  &  0.212161 e^{ -i0.000153} &  \\ 0.000000  &  0.000000  &  0.000000  &  0.018340  & 1.030141  &  \\ 0.000000  &  0.000000  &  0.143237 e^{ -i0.038278} &  0.846612  &  0.737932  \\ \end{pmatrix} \end{equation}

\begin{equation} Y_{d} = \begin{pmatrix}  0.004910 e^{ i0.542198} &  0.009242 e^{ i1.094004} & 0.000000  &  0.000000  &  0.000000  & \\ 0.043988 e^{ i0.860264} &  0.070637 e^{ i0.786049} &  0.000000  &  0.000000   &  0.000000  & \\ 0.000000  &  0.000000  &  0.862762 e^{ i0.789853} &  0.000000  &  0.507232 e^{ -i0.000123} &  \\ 0.000000  &  0.000000  &  0.000000  &  0.001187  & 0.056977  &  \\ 0.000000  &  0.000000  &  0.143237 e^{ -i0.038278} &  0.846612  &  0.501296  \\ \end{pmatrix} \end{equation}

\begin{equation} Y_{e} = \begin{pmatrix}  0.000039  &  0.000002  & 0.000000  &  0.000000  &  0.000000  & \\ 0.000103  &  -0.000084 &  0.000000  &  0.000000   &  0.000000  & \\ 0.000000  &  0.000000  &  0.472658 e^{ i0.789853} &  0.000000  &  0.000000  &  \\ 0.000000  &  0.000000  &  0.000000  &  0.028901  & 0.150963  &  \\ 0.000000  &  0.000000  &  0.000000  &  0.118708  &  0.497205  \\ \end{pmatrix} \end{equation}

\subsubsection{RH Neutrino mass parameters and the neutrino mass matrix at the weak scale}
\label{sec:7.5.2}
The mass parameters $M_L, M_R$ and $M_R^{ab}$ in the $\mathcal{L}_{maj}$ of Equation (2.3) are given by: \\
\begin{equation}
    M_L = M_R = 8.55285 \times 10^{12} \ GeV
\end{equation}
\begin{equation} M_R^{ab} = \begin{pmatrix}  1.450248 e^{ -i0.159487} \times 10^{11} &  3.369465 \times 10^{11} & \\ 3.369465 \times 10^{11}  &  4.117877 e^{ -i0.167092} \times 10^{11}  \\ \end{pmatrix} \end{equation}
These contribute to 6 additional parameters. \\
The neutrino mass matrix at the weak scale is given by:
\begin{equation} M_{neutrino} = \begin{pmatrix}  2.7816 e^{ i0.4396} \times 10^{-10} &  2.8478 e^{ i0.6337} \times 10^{-10} & 0.000000  &  0.000000  &  1300.727081  & \\ 2.8478 e^{ i0.6337} \times 10^{-10} &  2.952 e^{ i0.9654}  \times 10^{-10} &  0.000000  &  0.000000   &  1300.727081  & \\ 0.000000  &  0.000000  &  0.000000  &  4.5650 \times 10^{-11}  &  0.000000  &  \\ 0.000000  &  0.000000  &  4.5650 \times 10^{-11}  &  0.000000  & 361.055191  &  \\ 1300.727081  &  1300.727081  &  0.000000  &  361.055191  &  2.1814 \times 10^{-14}  \\ \end{pmatrix} \end{equation}
All the entries in $M_R^{ab}$ and $M_{neutrino}$ are in GeV.

\subsubsection{Soft SUSY breaking parameters}
\label{sec:7.5.3}
%\begin{comment}
The GMM parameters for this fit point are given by:
\begin{equation}
    l_a = 1, \ \alpha = 30, \ a_0 = 0, \ m_{3/2} = \ 20 \ TeV, \ c_m = 14, \ c_{H_u} = 19.81999, \ c_{H_d} = 12.79353
\end{equation}
Here, $m_{3/2}$ is the gravitino mass and we have taken a universal $a_{ijk} = a_0$. Hence, we have 6 soft SUSY breaking parameters. \\
The weak scale soft SUSY breaking mass matrices are given by:
\begin{equation} m^2_{\textbf{Q}} = \begin{pmatrix}  223.953696  &  6.047421 e^{ i3.135137} & 0.000000  &  0.000000  &  0.000000  & \\ 6.047421 e^{ -i3.135137} &  223.721080 &  0.000000  &  0.000000   &  0.000000  & \\ 0.000000  &  0.000000  &  110.803934  &  0.000000  &  0.000000  &  \\ 0.000000  &  0.000000  &  0.000000  &  151.289535  & 0.000000  &  \\ 0.000000  &  0.000000  &  0.000000  &  0.000000  &  94.077908  \\ \end{pmatrix} \end{equation}

\begin{equation} m^2_{\bar{\textbf{u}}} = \begin{pmatrix}  219.191451 &  11.548647 e^{ -i3.141563} & 0.000000  &  0.000000  &  0.000000  & \\ 11.548647 e^{ i3.141563} &  219.194739  &  0.000000  &  0.000000   &  0.000000  & \\ 0.000000  &  0.000000  &  78.855963 &  0.000000  &  0.000000  &  \\ 0.000000  &  0.000000  &  0.000000  &  28.764739  & 0.000000  &  \\ 0.000000  &  0.000000  &  0.000000  &  0.000000  &  99.357419  \\ \end{pmatrix} \end{equation}

\begin{equation} m^2_{\bar{\textbf{d}}} = \begin{pmatrix}  224.930034 &  0.325380 e^{ i3.074355} & 0.000000  &  0.000000  &  0.000000  & \\ 0.325380 e^{ -i3.074355} &  224.164090  &  0.000000  &  0.000000   &  0.000000  & \\ 0.000000  &  0.000000  &  50.041348 &  0.000000  &  0.000000  &  \\ 0.000000  &  0.000000  &  0.000000  &  133.993569  & 0.000000  &  \\ 0.000000  &  0.000000  &  0.000000  &  0.000000  &  209.549257  \\ \end{pmatrix} \end{equation}

\begin{equation} m^2_{\textbf{L}} = \begin{pmatrix}  177.008701 e^{ i-0.000000} &  51.893498 e^{ -i3.141511} & 0.000000  &  0.000000  &  0.000000  & \\ 51.893498 e^{ i3.141511} &  171.787289 &  0.000000  &  0.000000   &  0.000000  & \\ 0.000000  &  0.000000  &  122.499993  &  0.000000  &  0.000000  &  \\ 0.000000  &  0.000000  &  0.000000  &  150.724394  & 0.000000  &  \\ 0.000000  &  0.000000  &  0.000000  &  0.000000  &  90.356545  \\ \end{pmatrix} \end{equation}

\begin{equation} m^2_{\bar{\textbf{e}}} = \begin{pmatrix}  197.468017  &  1.279482 e^{ i3.141585} & 0.000000  &  0.000000  &  0.000000  & \\ 1.279482 e^{ -i3.141585} &  197.346591 &  0.000000  &  0.000000   &  0.000000  & \\ 0.000000  &  0.000000  &  95.752434 &  0.000000  &  0.000000  &  \\ 0.000000  &  0.000000  &  0.000000  &  63.090522  & 0.000000  &  \\ 0.000000  &  0.000000  &  0.000000  &  0.000000  &  126.769184  \\ \end{pmatrix} \end{equation}

\begin{equation} m^2_{\Bar{\textbf{$\nu$}}} = \begin{pmatrix}  119.902142 &  100.622647 e^{ -i3.141512} & 0.000000  &  0.000000  &  0.000000  & \\ 100.622647 e^{ i3.141512} &  148.270196 &  0.000000  &  0.000000   &  0.000000  & \\ 0.000000  &  0.000000  &  182.846500 &  0.000000  &  0.000000  &  \\ 0.000000  &  0.000000  &  0.000000  &  77.310330  & 0.000000  &  \\ 0.000000  &  0.000000  &  0.000000  &  0.000000  &  5.566623  \\ \end{pmatrix} \end{equation}
All the entries in the above matrices are in TeV$^2$. \\
The weak scale gaugino mass parameters $M_a$ are given by:
\begin{equation}
    M_1 = 628.093 \ GeV, \ M_2 = 992.049 \ GeV, \ M_3 = 2.28438 \ TeV, \ M_4 = 429.250 \ GeV
\end{equation}

\subsection{Extended CKM matrix at the fit point}
\label{sec:7.6}

\begin{equation} \hat{V}_{CKM} = \begin{pmatrix}  0.973997 e^{ i0.137293} &  0.226529 e^{ -i1.475268} & 0.003769 e^{ i0.308875} &  0.000001 e^{ i2.610939} &  0.000006 e^{ i2.611603} & \\ 0.226416 e^{ i1.982142} &  0.973174 e^{ -i2.772616} &  0.040843 e^{ i0.060402} &  0.000006 e^{ i2.368311}  &  0.000065 e^{ i2.322401} & \\ 0.008076 e^{ i1.505322} &  0.040214 e^{ -i2.825819} &  0.999053 e^{ i3.130504} &  0.000381 e^{ i3.102532} &  0.000352 e^{ i2.245719} &  \\ 0.000045 e^{ -i2.917042} &  0.000432 e^{ -i1.357450} &  0.014505 e^{ -i2.697973} &  0.023191 e^{ i3.126263} & 0.001407 e^{ -i0.013187} &  \\ 0.000003 e^{ i1.922230} &  0.000021 e^{ -i2.239466} &  0.000684 e^{ -i2.930490} &  0.997883 e^{ i0.000004} &  0.060737 e^{ -i3.140826} \\ \end{pmatrix} \end{equation}
%\end{comment}

\subsection{$Z'$ couplings at the fit point}
\label{sec:7.7}

\begin{equation} \hat{g}^u_L = \begin{pmatrix}  -0.000000434 &  0.000319 e^{ -i3.141175} & 0.000069 e^{ -i3.078618} &  0.000014 e^{ i3.123712} &  0.000572 e^{ -i0.000438} & \\ 0.000319 e^{ i3.141175} &  -0.234369 &  0.050764 e^{ -i3.084329} &  0.010336 e^{ i3.123154}  &  0.420433 e^{ -i0.000778} & \\ 0.000069 e^{ i3.078618} &  0.050764 e^{ i3.084329} &  0.975915 &  0.010109 e^{ -i0.656617} &  0.204143 e^{ -i0.026670} &  \\ 0.000014 e^{ -i3.123712} &  0.010336 e^{ -i3.123154} &  0.010109 e^{ i0.656617} &  -0.000309 & 0.019742 e^{ i0.053349} &  \\ 0.000572 e^{ i0.000438} &  0.420433 e^{ i0.000778} &  0.204143 e^{ i0.026670} &  0.019742 e^{ -i0.053349} &  -0.741236 \\ \end{pmatrix} \end{equation}

\begin{equation} \hat{g}^u_R = \begin{pmatrix}  -0.000001 &  0.000489 e^{ i3.141260} & 0.000260 e^{ i3.126864} &  0.001072 e^{ i0.000509} &  0.000024 e^{ i3.124938} & \\ 0.000489 e^{ -i3.141260} &  -0.164348 &  0.087195 e^{ i3.129088} &  0.360096 e^{ i0.000731}  &  0.008174 e^{ i3.125094} & \\ 0.000260 e^{ -i3.126864} &  0.087195 e^{ -i3.129088} &  0.899619 &  0.417027 e^{ i0.005909} &  0.015488 e^{ -i2.956363} &  \\ 0.001072 e^{ -i0.000509} &  0.360096 e^{ -i0.000731} &  0.417027 e^{ -i0.005909} &  -0.734997  & 0.015339 e^{ -i0.064909} &  \\ 0.000024 e^{ -i3.124938} &  0.008174 e^{ -i3.125094} &  0.015488 e^{ i2.956363} &  0.015339 e^{ i0.064909} &  -0.000273 \\ \end{pmatrix} \end{equation}

\begin{equation} \hat{g}^d_L = \begin{pmatrix}  -0.012083 &  0.052099 e^{ -i1.606585} & 0.003798 e^{ -i2.934006} &  0.096382 e^{ -i1.969962} &  0.005864 e^{ i1.173160} & \\ 0.052099 e^{ i1.606585} &  -0.224384 &  0.000595 e^{ i2.864449} &  0.416731 e^{ i2.772071}  &  0.025352 e^{ -i0.368221} & \\ 0.003798 e^{ i2.934006} &  0.000595 e^{ -i2.864449} &  0.978368 &  0.185601 e^{ i3.130980} &  0.011541 e^{ -i0.032619} &  \\ 0.096382 e^{ i1.969962} &  0.416731 e^{ -i2.772071} &  0.185601 e^{ -i3.130980} &  -0.739172 & 0.044920 e^{ i0.002430} &  \\ 0.005864 e^{ -i1.173160} &  0.025352 e^{ i0.368221} &  0.011541 e^{ i0.032619} &  0.044920 e^{ -i0.002430} &  -0.002730 \\ \end{pmatrix} \end{equation}

\begin{equation} \hat{g}^d_R = \begin{pmatrix}  -0.945118 &  0.224121 e^{ -i0.001067} & 0.017158 e^{ -i3.135415} &  0.002714 e^{ i0.020878} &  0.036574 e^{ -i3.120954} & \\ 0.224121 e^{ i0.001067} &  -0.053144 &  0.002537 e^{ i0.074178} &  0.000697 e^{ -i3.129376}  &  0.009398 e^{ i0.012179} & \\ 0.017158 e^{ i3.135415} &  0.002537 e^{ -i0.074178} &  0.816818 &  0.028594 e^{ -i0.001210} &  0.386166 e^{ -i3.140173} &  \\ 0.002714 e^{ -i0.020878} &  0.000697 e^{ i3.129376} &  0.028594 e^{ i0.001210} &  0.000989 & 0.013362 e^{ -i3.138935} &  \\ 0.036574 e^{ i3.120954} &  0.009398 e^{ -i0.012179} &  0.386166 e^{ i3.140173} &  0.013362 e^{ i3.138935} &  0.180455 \\ \end{pmatrix} \end{equation}

\begin{equation} \hat{g}^e_L = \begin{pmatrix}  -0.0000000322 &  0.000176 e^{ i3.141593} & 0.000000 &  0.000024  &  -0.000024 & \\ -0.000176 &  -0.962904 &  0.000000 &  0.133323   &  -0.133958 & \\ 0.000000 &  0.000000 &  1.000000  &  0.000000  &  0.000000 &  \\ 0.000024  &  0.133323  &  0.000000  &  -0.018460 & 0.018548  &  \\ -0.000024 &  -0.133958 &  0.000000 &  0.018548  &  -0.018636 \\ \end{pmatrix} \end{equation}

\begin{equation} \hat{g}^e_R = \begin{pmatrix}  -0.0000000079 &  -0.000087 & 0.000000  &  0.000012 &  -0.000012 & \\ -0.000087 &  -0.962814 &  0.000000 &  0.137147  &  -0.130362 & \\ 0.000000 &  0.000000 &  1.000000  &  0.000000  &  0.000000 &  \\ 0.000012  &  0.137147  &  0.000000  &  -0.019536  & 0.018569  &  \\ -0.000012 &  -0.130362 &  0.000000 &  0.018569  &  -0.017651 \\ \end{pmatrix} \end{equation}

\subsection{Input parameters at the fit point - Fit point II}
\label{sec:7.8}
\begin{equation}
    g_1 = g_2 = g_3 = g_4 = g_{GUT} = 1.4206, \ \epsilon_3 = 0, \ M_{GUT} = 5.012 \times 10^{16} \ GeV
\end{equation}
\begin{equation}
    tan\beta = 47.62315, \ v_\Phi = 3039.259 \ GeV, \ v_{\bar \Phi} = 2388.616 \ GeV
\end{equation}
\begin{equation}
    \lambda_{2Q_1} = 838.027 \ GeV, \ \lambda_{2Q_2} = 1147.758 \ GeV, \ \lambda_{2L_1} = \lambda_{2L_2} = 1290.611 \ GeV
\end{equation}
\begin{equation}
    \lambda_{2U_1} = 646.556 \ GeV, \ \lambda_{2U_2} = 881.232 \ GeV, \ \lambda_{2D_1} = 989.959 \ GeV, \lambda_{2D_2} = 3100.593 \ GeV
\end{equation}
\begin{equation}
    \lambda_{2E_1} = -1823.503 \ GeV, \ \lambda_{2E_2} = 0 \ GeV
\end{equation}
The remaining 50 input parameters are presented in the form of the Yukawa matrices at the GUT scale and the soft SUSY breaking parameters. Many of the entries in the Yukawa matrices are 0 and some of them are equal (in the GUT sense) which together give 38 of the model parameters. The Yukawa matrices at the GUT scale are:
\begin{equation} Y_{u} = \begin{pmatrix}  0.000009357  &  0.000000  & 0.000000  &  0.000000  &  0.000000  & \\ 0.000000  &  0.000000  &  0.000000  &  0.000000   &  0.000000  & \\ 0.000000  &  0.000000  &  1.827175 e^{ i0.790233} &  0.000000  &  1.188228 e^{ -i0.000217} &  \\ 0.000000  &  0.000000  &  0.000000  &  0.017894  & 1.618202  &  \\ 0.000000  &  0.000000  &  0.152240 e^{ -i0.043436} &  2.194159  &  1.963979  \\ \end{pmatrix} \end{equation}

\begin{equation} Y_{d} = \begin{pmatrix}  0.002409  &  0.006097 e^{ i1.192444} & 0.000000  &  0.000000  &  0.000000  & \\ 0.016879 e^{ i0.925307} &  0.044389 e^{ i0.785451} &  0.000000  &  0.000000   &  0.000000  & \\ 0.000000  &  0.000000  &  1.827175 e^{ i0.790233} &  0.000000  &  2.381021 e^{ -i0.000132} &  \\ 0.000000  &  0.000000  &  0.000000  &  0.001168  & 0.040008  &  \\ 0.000000  &  0.000000  &  0.152240 e^{ -i0.043436} &  2.194159  &  0.795780  \\ \end{pmatrix} \end{equation}

\begin{equation} Y_{e} = \begin{pmatrix}  0.000204  &  0.000184  & 0.000000  &  0.000000  &  0.000000  & \\ 0.000186  &  0.000000  &  0.000000  &  0.000000   &  0.000000  & \\ 0.000000  &  0.000000  &  1.827175 e^{ i0.790233} &  0.000000  &  0.000000  &  \\ 0.000000  &  0.000000  &  0.000000  &  0.073570  & 0.472727  &  \\ 0.000000  &  0.000000  &  0.000000  &  0.861073  &  2.959575  \\ \end{pmatrix} \end{equation}

\begin{equation} Y_{\nu} = \begin{pmatrix}  2.191883  &  2.316690  & 0.000000  &  0.000000  &  0.000000  & \\ 1.825601  &  1.838974  &  0.000000  &  0.000000   &  0.000000  & \\ 0.000000  &  0.000000  &  1.381811  &  0.000000  &  0.000000  &  \\ 0.000000  &  0.000000  &  0.000000  &  1.785259  & 1.947156  &  \\ 0.000000  &  0.000000  &  0.000000  &  0.861073  &  2.070352  \\ \end{pmatrix} \end{equation}

\subsubsection{Quark and charged lepton Yukawa matrices at the weak scale}

\begin{equation} Y_{u} = \begin{pmatrix}  0.00001546 e^{ i0.000001} &  0.000006 e^{ i0.000060} & 0.000000  &  0.000000  &  0.000000  & \\ 0.000000  &  0.000000  &  0.000000  &  0.000000   &  0.000000  & \\ 0.000000  &  0.000000  &  0.989540 e^{ i0.790233} &  0.000000  &  0.167705 e^{ -i0.000217} &  \\ 0.000000  &  0.000000  &  0.000000  &  0.018252  & 1.024819  &  \\ 0.000000  &  0.000000  &  0.140206 e^{ -i0.043436} &  0.848179  &  0.723105  \\ \end{pmatrix} \end{equation}

\begin{equation} Y_{d} = \begin{pmatrix}  0.005125 e^{ i0.510512} &  0.009232 e^{ i1.085049} & 0.000000  &  0.000000  &  0.000000  & \\ 0.043864 e^{ i0.862817} &  0.071353 e^{ i0.800050} &  0.000000  &  0.000000   &  0.000000  & \\ 0.000000  &  0.000000  &  0.858954 e^{ i0.790233} &  0.000000  &  0.493623 e^{ -i0.000132} &  \\ 0.000000  &  0.000000  &  0.000000  &  0.001492  & 0.053696  &  \\ 0.000000  &  0.000000  &  0.140206 e^{ -i0.043436} &  0.848179  &  0.495800  \\ \end{pmatrix} \end{equation}

\begin{equation} Y_{e} = \begin{pmatrix}  0.000032  &  0.000011  & 0.000000  &  0.000000  &  0.000000  & \\ 0.000104  &  -0.000083 &  0.000000  &  0.000000   &  0.000000  & \\ 0.000000  &  0.000000  &  0.472927 e^{ i0.790233} &  0.000000  &  0.000000  &  \\ 0.000000  &  0.000000  &  0.000000  &  0.028925  & 0.149592  &  \\ 0.000000  &  0.000000  &  0.000000  &  0.118978  &  0.504921  \\ \end{pmatrix} \end{equation}

\subsubsection{RH Neutrino mass parameters and the neutrino mass matrix at the weak scale}
\label{sec:7.8.2}
The mass parameters $M_L, M_R$ and $M_R^{ab}$ in the $\mathcal{L}_{maj}$ of Equation (2.3) are given by: \\
\begin{equation}
    M_L = M_R = 1.2504 \times 10^{13} \ GeV
\end{equation}
\begin{equation} M_R^{ab} = \begin{pmatrix}  1.812229 e^{ -i0.268412} \times 10^{11} &  4.947073 \times 10^{11} & \\ 4.947073 \times 10^{11} &  5.983751 e^{ -i0.263383} \times 10^{11}  \\ \end{pmatrix} \end{equation}
These contribute to 6 additional parameters. \\
The neutrino mass matrix at the weak scale is given by:
\begin{equation} M_{neutrino} = \begin{pmatrix}  2.5793 e^{ i0.4362} \times 10^{-10} &  2.6209 e^{ i0.6416} \times 10^{-10} & 0.000000  &  0.000000  &  1290.610742  & \\ 2.6209 e^{ i0.6416} \times 10^{-10} &  2.7256 e^{ i1.0088} \times 10^{-10} &  0.000000  &  0.000000   &  1290.610742  & \\ 0.000000  &  0.000000  &  0.000000  &  4.6185 \times 10^{-11}  &  0.000000  &  \\ 0.000000  &  0.000000  &  4.6185 \times 10^{-11}  &  0.000000  & 361.606446  &  \\ 1290.610742  &  1290.610742  &  0.000000  &  361.606446  &  1.6712 \times 10^{-14}  \\ \end{pmatrix} \end{equation}
All the entries in $M_R^{ab}$ and $M_{neutrino}$ are in GeV.

\subsubsection{Soft SUSY breaking parameters}
\label{sec:7.8.3}
%\begin{comment}
The GMM parameters for this fit point are given by:
\begin{equation}
    l_a = 1, \ \alpha = 30, \ a_0 = 0, \ m_{3/2} = \ 25 \ TeV, \ c_m = 23, \ c_{H_u} = 19.42162, \ c_{H_d} = 36.76136
\end{equation}
Here, $m_{3/2}$ is the gravitino mass and we have taken a universal $a_{ijk} = a_0$. Hence, we have 6 soft SUSY breaking parameters. \\
The weak scale soft SUSY breaking mass matrices are given by:
\begin{equation} m^2_{\textbf{Q}} = \begin{pmatrix}  556.868567 &  8.408083 e^{ i3.133000} & 0.000000  &  0.000000  &  0.000000  & \\ 8.408083 e^{ -i3.133000} &  556.200694 &  0.000000  &  0.000000   &  0.000000  & \\ 0.000000  &  0.000000  &  269.643182 &  0.000000  &  0.000000  &  \\ 0.000000  &  0.000000  &  0.000000  &  395.189964  & 0.000000  &  \\ 0.000000  &  0.000000  &  0.000000  &  0.000000  &  221.828595  \\ \end{pmatrix} \end{equation}

\begin{equation} m^2_{\Bar{\textbf{u}}} = \begin{pmatrix}  532.341519  &  15.608831 e^{ -i3.141568} & 0.000000  &  0.000000  &  0.000000  & \\ 15.608831 e^{ i3.141568} &  532.345994  &  0.000000  &  0.000000   &  0.000000  & \\ 0.000000  &  0.000000  &  270.150118  &  0.000000  &  0.000000  &  \\ 0.000000  &  0.000000  &  0.000000  &  96.193666  & 0.000000  &  \\ 0.000000  &  0.000000  &  0.000000  &  0.000000  &  293.824856  \\ \end{pmatrix} \end{equation}

\begin{equation} m^2_{\Bar{\textbf{d}}} = \begin{pmatrix}  560.954916  &  0.625278 e^{ i3.116404} & 0.000000  &  0.000000  &  0.000000  & \\ 0.625278 e^{ -i3.116404} &  558.846834 &  0.000000  &  0.000000   &  0.000000  & \\ 0.000000  &  0.000000  &  122.070633 &  0.000000  &  0.000000  &  \\ 0.000000  &  0.000000  &  0.000000  &  369.859074  & 0.000000  &  \\ 0.000000  &  0.000000  &  0.000000  &  0.000000  &  530.700060  \\ \end{pmatrix} \end{equation}

\begin{equation} m^2_{\textbf{L}} = \begin{pmatrix}  441.833308 &  106.881354 e^{ -i3.141549} & 0.000000  &  0.000000  &  0.000000  & \\ 106.881354 e^{ i3.141549} &  432.057854 &  0.000000  &  0.000000   &  0.000000  & \\ 0.000000  &  0.000000  &  259.406990  &  0.000000  &  0.000000  &  \\ 0.000000  &  0.000000  &  0.000000  &  429.168777  & 0.000000  &  \\ 0.000000  &  0.000000  &  0.000000  &  0.000000  &  250.349982  \\ \end{pmatrix} \end{equation}

\begin{equation} m^2_{\Bar{\textbf{e}}} = \begin{pmatrix}  533.723453 &  1.918893 e^{ i3.141586} & 0.000000  &  0.000000  &  0.000000  & \\ 1.918893 e^{ -i3.141586} &  533.557031  &  0.000000  &  0.000000   &  0.000000  & \\ 0.000000  &  0.000000  &  240.370308  &  0.000000  &  0.000000  &  \\ 0.000000  &  0.000000  &  0.000000  &  205.618206  & 0.000000  &  \\ 0.000000  &  0.000000  &  0.000000  &  0.000000  &  452.394311  \\ \end{pmatrix} \end{equation}

\begin{equation} m^2_{\Bar{\textbf{$\nu$}}} = \begin{pmatrix}  321.872232 &  207.072839 e^{ -i3.141549} & 0.000000  &  0.000000  &  0.000000  & \\ 207.072839 e^{ i3.141549} &  393.081059 &  0.000000  &  0.000000   &  0.000000  & \\ 0.000000  &  0.000000  &  402.031334 &  0.000000  &  0.000000  &  \\ 0.000000  &  0.000000  &  0.000000  &  152.922796  & 0.000000  &  \\ 0.000000  &  0.000000  &  0.000000  &  0.000000  &  168.027535  \\ \end{pmatrix} \end{equation}
All the entries in the above matrices are in TeV$^2$. \\
The weak scale gaugino mass parameters $M_a$ are given by:
\begin{equation}
    M_1 = 811.985 \ GeV, \ M_2 = 1.27978 \ TeV, \ M_3 = 2.91131 \ TeV, \ M_4 = 544.317 \ GeV
\end{equation}

\subsection{Extended CKM matrix at the fit point}
\label{sec:7.9}

\begin{equation} \hat{V}_{CKM} = \begin{pmatrix}  0.973993 e^{ i0.134100} &  0.226545 e^{ -i1.517993} & 0.003769 e^{ i0.230614} &  0.000001 e^{ i2.529773} &  0.000006 e^{ i2.530413} & \\ 0.226437 e^{ i2.024459} &  0.973206 e^{ -i2.769825} &  0.039961 e^{ i0.048011} &  0.000006 e^{ i2.358623}  &  0.000066 e^{ i2.315121} & \\ 0.007968 e^{ i1.550805} &  0.039341 e^{ -i2.808380} &  0.999126 e^{ i3.132113} &  0.000376 e^{ i3.096810} &  0.000345 e^{ i2.243055} &  \\ 0.000048 e^{ -i2.313704} &  0.000388 e^{ -i1.024620} &  0.011685 e^{ -i2.644382} &  0.016906 e^{ i3.123424} & 0.001051 e^{ -i0.016141} &  \\ 0.000003 e^{ i1.870152} &  0.000016 e^{ -i2.303574} &  0.000542 e^{ -i2.967029} &  0.997915 e^{ i0.000001} &  0.062285 e^{ -i3.140812} \\ \end{pmatrix} \end{equation}
%\end{comment}

\subsection{$Z'$ couplings at the fit point}
\label{sec:7.10}

\begin{equation} \hat{g}^u_L = \begin{pmatrix}  -0.000000418  &  0.000312 e^{ -i3.141265} & 0.000066 e^{ -i3.089746} &  0.000010 e^{ i3.119136} &  0.000562 e^{ -i0.000345} & \\ 0.000312 e^{ i3.141265} &  -0.232980 &  0.049228 e^{ -i3.094530} &  0.007512 e^{ i3.118696}  &  0.419792 e^{ -i0.000611} & \\ 0.000066 e^{ i3.089746} &  0.049228 e^{ i3.094530} &  0.977043 &  0.008631 e^{ -i0.695958} &  0.199636 e^{ -i0.021708} &  \\ 0.000010 e^{ -i3.119136} &  0.007512 e^{ -i3.118696} &  0.008631 e^{ i0.695958} &  -0.000142 & 0.014486 e^{ i0.064603} &  \\ 0.000562 e^{ i0.000345} &  0.419792 e^{ i0.000611} &  0.199636 e^{ i0.021708} &  0.014486 e^{ -i0.064603} &  -0.743921 \\ \end{pmatrix} \end{equation}

\begin{equation} \hat{g}^u_R = \begin{pmatrix}  -0.000001265 &  0.000458 e^{ i3.141333} & 0.000195 e^{ i3.123410} &  0.001008 e^{ i0.000400} &  0.000017 e^{ i3.121793} & \\ 0.000458 e^{ -i3.141333} &  -0.166148 &  0.070369 e^{ i3.126002} &  0.365452 e^{ i0.000573}  &  0.006057 e^{ i3.121893} & \\ 0.000195 e^{ -i3.123410} &  0.070369 e^{ -i3.126002} &  0.935430  &  0.337750 e^{ i0.007258} &  0.011516 e^{ -i2.865916} &  \\ 0.001008 e^{ -i0.000400} &  0.365452 e^{ -i0.000573} &  0.337750 e^{ -i0.007258} &  -0.769145 & 0.011738 e^{ -i0.073905} &  \\ 0.000017 e^{ -i3.121793} &  0.006057 e^{ -i3.121893} &  0.011516 e^{ i2.865916} &  0.011738 e^{ i0.073905} &  -0.000135 \\ \end{pmatrix} \end{equation}

\begin{equation} \hat{g}^d_L = \begin{pmatrix}  -0.012002  &  0.051759 e^{ -i1.646304} & 0.003709 e^{ -i3.025590} &  0.096148 e^{ -i2.012485} &  0.005999 e^{ i1.130630} & \\ 0.051759 e^{ i1.646304} &  -0.222950 &  0.000290 e^{ -i2.555244} &  0.415680 e^{ i2.769248}  &  0.025932 e^{ -i0.371040} & \\ 0.003709 e^{ i3.025590} &  0.000290 e^{ i2.555244} &  0.979294 &  0.181588 e^{ i3.133261} &  0.011572 e^{ -i0.029921} &  \\ 0.096148 e^{ i2.012485} &  0.415680 e^{ -i2.769248} &  0.181588 e^{ -i3.133261} &  -0.741462 & 0.046211 e^{ i0.002368} &  \\ 0.005999 e^{ -i1.130630} &  0.025932 e^{ i0.371040} &  0.011572 e^{ i0.029921} &  0.046211 e^{ -i0.002368} &  -0.002880 \\ \end{pmatrix} \end{equation}

\begin{equation} \hat{g}^d_R = \begin{pmatrix}  -0.947128 &  0.220492 e^{ -i0.003894} & 0.015890 e^{ -i3.088481} &  0.002621 e^{ i0.072084} &  0.034653 e^{ -i3.069740} & \\ 0.220492 e^{ i0.003894} &  -0.051328 &  0.002191 e^{ i0.153794} &  0.000663 e^{ -i3.078312}  &  0.008763 e^{ i0.063249} & \\ 0.015890 e^{ i3.088481} &  0.002191 e^{ -i0.153794} &  0.824729  &  0.028649 e^{ -i0.001440} &  0.379471 e^{ -i3.140449} &  \\ 0.002621 e^{ -i0.072084} &  0.000663 e^{ i3.078312} &  0.028649 e^{ i0.001440} &  0.000985 & 0.013041 e^{ -i3.138984} &  \\ 0.034653 e^{ i3.069740} &  0.008763 e^{ -i0.063249} &  0.379471 e^{ i3.140449} &  0.013041 e^{ i3.138984} &  0.172742 \\ \end{pmatrix} \end{equation}

\begin{equation} \hat{g}^e_L = \begin{pmatrix}  -0.00000001 &  -0.000098 & 0.000000  &  0.000014  &  -0.000014 & \\ -0.000098 &  -0.962232 &  0.000000  &  0.134474   &  -0.135125  & \\ 0.000000  &  0.000000  &  1.000000  &  0.000000 &  0.000000  &  \\ 0.000014  &  0.134474  &  0.000000 &  -0.018793  & 0.018884  &  \\ -0.000014 &  -0.135125 &  0.000000  &  0.018884  &  -0.018976 \\ \end{pmatrix} \end{equation}

\begin{equation} \hat{g}^e_R = \begin{pmatrix}  -0.00000001 &  -0.000099 & 0.000000 &  0.000014  &  -0.000013 & \\ -0.000099  &  -0.963023 &  0.000000 &  0.136850   &  -0.129930 & \\ 0.000000  &  0.000000 &  1.000000  &  0.000000  &  0.000000  &  \\ 0.000014  &  0.136850  &  0.000000  &  -0.019447  & 0.018464  &  \\ -0.000013 &  -0.129930 &  0.000000 &  0.018464  &  -0.017530 \\ \end{pmatrix} \end{equation}

\section{ACKNOWLEDGEMENTS}
We would like to thank Junichiro Kawamura for reading over the first draft and making some important comments. The work of S.R. is supported in part by the Department of Energy (DOE) under Award No. DE-SC0011726. The work of H.K. is supported by the KVPY fellowship of the Department of Science and Technology (DST), Government of India.

% The bibliography will probably be heavily edited during typesetting.
% We'll parse it and, using the arxiv number or the journal data, will
% query inspire, trying to verify the data (this will probalby spot
% eventual typos) and retrive the document DOI and eventual errata.
% We however suggest to always provide author, title and journal data:
% in short all the informations that clearly identify a document.

\bibliographystyle{unsrt}
\bibliography{references}

\begin{thebibliography}{10}

\bibitem{Muong-2:2006rrc}
G.~W. Bennett et~al.
\newblock {Final Report of the Muon E821 Anomalous Magnetic Moment Measurement
  at BNL}.
\newblock {\em Phys. Rev. D}, 73:072003, 2006.

\bibitem{Muong-2:2021ojo}
B.~Abi et~al.
\newblock {Measurement of the Positive Muon Anomalous Magnetic Moment to 0.46
  ppm}.
\newblock {\em Phys. Rev. Lett.}, 126(14):141801, 2021.

\bibitem{Aoyama:2020ynm}
T.~Aoyama et~al.
\newblock {The anomalous magnetic moment of the muon in the Standard Model}.
\newblock {\em Phys. Rept.}, 887:1--166, 2020.

\bibitem{LHCb:2014cxe}
R.~Aaij et~al.
\newblock {Differential branching fractions and isospin asymmetries of $B \to
  K^{(*)} \mu^+ \mu^-$ decays}.
\newblock {\em JHEP}, 06:133, 2014.

\bibitem{LHCb:2014vgu}
Roel Aaij et~al.
\newblock {Test of lepton universality using $B^{+}\rightarrow
  K^{+}\ell^{+}\ell^{-}$ decays}.
\newblock {\em Phys. Rev. Lett.}, 113:151601, 2014.

\bibitem{LHCb:2013tgx}
R~Aaij et~al.
\newblock {Differential branching fraction and angular analysis of the decay
  $B_s^0\to\phi\mu^{+}\mu^{-}$}.
\newblock {\em JHEP}, 07:084, 2013.

\bibitem{BaBar:2013qry}
J.~P. Lees et~al.
\newblock {Measurement of the $B \to X_s l^+l^-$ branching fraction and search
  for direct CP violation from a sum of exclusive final states}.
\newblock {\em Phys. Rev. Lett.}, 112:211802, 2014.

\bibitem{LHCb:2015wdu}
Roel Aaij et~al.
\newblock {Angular analysis and differential branching fraction of the decay
  $B^0_s\to\phi\mu^+\mu^-$}.
\newblock {\em JHEP}, 09:179, 2015.

\bibitem{LHCb:2013ghj}
R~Aaij et~al.
\newblock {Measurement of Form-Factor-Independent Observables in the Decay
  $B^{0} \to K^{*0} \mu^+ \mu^-$}.
\newblock {\em Phys. Rev. Lett.}, 111:191801, 2013.

\bibitem{LHCb:2015svh}
Roel Aaij et~al.
\newblock {Angular analysis of the $B^{0} \to K^{*0} \mu^{+} \mu^{-}$ decay
  using 3 fb$^{-1}$ of integrated luminosity}.
\newblock {\em JHEP}, 02:104, 2016.

\bibitem{CMS:2017ivg}
{Measurement of the $P_1$ and $P_5'$ angular parameters of the decay
  $\mathrm{B}^0 \to \mathrm{K}^{*0} \mu^+ \mu^-$ in proton-proton collisions at
  $\sqrt{s}=8~\mathrm{TeV}$}.
\newblock 2017.

\bibitem{CMS:2015bcy}
Vardan Khachatryan et~al.
\newblock {Angular analysis of the decay $B^0 \to K^{*0} \mu^+ \mu^-$ from pp
  collisions at $\sqrt s = 8$ TeV}.
\newblock {\em Phys. Lett. B}, 753:424--448, 2016.

\bibitem{Belle:2016xuo}
A.~Abdesselam et~al.
\newblock {Angular analysis of $B^0 \to K^\ast(892)^0 \ell^+ \ell^-$}.
\newblock In {\em {LHC Ski 2016}: {A First Discussion of 13 TeV Results}}, 4
  2016.

\bibitem{Belle:2016fev}
S.~Wehle et~al.
\newblock {Lepton-Flavor-Dependent Angular Analysis of $B\to K^\ast
  \ell^+\ell^-$}.
\newblock {\em Phys. Rev. Lett.}, 118(11):111801, 2017.

\bibitem{ATLAS:2018gqc}
Morad Aaboud et~al.
\newblock {Angular analysis of $B^0_d \rightarrow K^{*}\mu^+\mu^-$ decays in
  $pp$ collisions at $\sqrt{s}= 8$ TeV with the ATLAS detector}.
\newblock {\em JHEP}, 10:047, 2018.

\bibitem{LHCb:2022zom}
{Measurement of lepton universality parameters in $B^+\to K^+\ell^+\ell^-$ and
  $B^0\to K^{*0}\ell^+\ell^-$ decays}.
\newblock 12 2022.

\bibitem{Czarnecki:2001pv}
Andrzej Czarnecki and William~J. Marciano.
\newblock {The Muon anomalous magnetic moment: A Harbinger for 'new physics'}.
\newblock {\em Phys. Rev. D}, 64:013014, 2001.

\bibitem{Kannike:2011ng}
Kristjan Kannike, Martti Raidal, David~M. Straub, and Alessandro Strumia.
\newblock {Anthropic solution to the magnetic muon anomaly: the charged
  see-saw}.
\newblock {\em JHEP}, 02:106, 2012.
\newblock [Erratum: JHEP 10, 136 (2012)].

\bibitem{Dermisek:2013gta}
Radovan Dermisek and Aditi Raval.
\newblock {Explanation of the Muon g-2 Anomaly with Vectorlike Leptons and its
  Implications for Higgs Decays}.
\newblock {\em Phys. Rev. D}, 88:013017, 2013.

\bibitem{Allanach:2015gkd}
Ben Allanach, Farinaldo~S. Queiroz, Alessandro Strumia, and Sichun Sun.
\newblock {$Z'$ models for the LHCb and $g-2$ muon anomalies}.
\newblock {\em Phys. Rev. D}, 93(5):055045, 2016.
\newblock [Erratum: Phys.Rev.D 95, 119902 (2017)].

\bibitem{Altmannshofer:2016oaq}
Wolfgang Altmannshofer, Marcela Carena, and Andreas Crivellin.
\newblock {$L_\mu - L_\tau$ theory of Higgs flavor violation and $(g-2)_\mu$}.
\newblock {\em Phys. Rev. D}, 94(9):095026, 2016.

\bibitem{Megias:2017dzd}
Eugenio Megias, Mariano Quiros, and Lindber Salas.
\newblock {$g_\mu-2$ from Vector-Like Leptons in Warped Space}.
\newblock {\em JHEP}, 05:016, 2017.

\bibitem{Buras:1994dj}
Andrzej~J. Buras and Manfred Munz.
\newblock {Effective Hamiltonian for B ---\ensuremath{>} X(s) e+ e- beyond
  leading logarithms in the NDR and HV schemes}.
\newblock {\em Phys. Rev. D}, 52:186--195, 1995.

\bibitem{Bobeth:1999mk}
Christoph Bobeth, Mikolaj Misiak, and Jorg Urban.
\newblock {Photonic penguins at two loops and $m_t$ dependence of $BR[B \to X_s
  l^+ l^-]$}.
\newblock {\em Nucl. Phys. B}, 574:291--330, 2000.

\bibitem{Raby:2017igl}
Stuart Raby and Andreas Trautner.
\newblock {Vectorlike chiral fourth family to explain muon anomalies}.
\newblock {\em Phys. Rev. D}, 97(9):095006, 2018.

\bibitem{CDF:2022hxs}
T.~Aaltonen et~al.
\newblock {High-precision measurement of the $W$ boson mass with the CDF II
  detector}.
\newblock {\em Science}, 376(6589):170--176, 2022.

\bibitem{ATLAS:2023fsi}
{Improved W boson Mass Measurement using 7 TeV Proton-Proton Collisions with
  the ATLAS Detector}.
\newblock 2023.

\bibitem{Kawamura:2022fhm}
Junichiro Kawamura and Stuart Raby.
\newblock {W mass in a model with vectorlike leptons and U(1)'}.
\newblock {\em Phys. Rev. D}, 106(3):035009, 2022.

\bibitem{Buchmuller:2005jr}
Wilfried Buchmuller, Koichi Hamaguchi, Oleg Lebedev, and Michael Ratz.
\newblock {Supersymmetric standard model from the heterotic string}.
\newblock {\em Phys. Rev. Lett.}, 96:121602, 2006.

\bibitem{Buchmuller:2006ik}
Wilfried Buchmuller, Koichi Hamaguchi, Oleg Lebedev, and Michael Ratz.
\newblock {Supersymmetric Standard Model from the Heterotic String (II)}.
\newblock {\em Nucl. Phys. B}, 785:149--209, 2007.

\bibitem{Lebedev:2006kn}
Oleg Lebedev, Hans~Peter Nilles, Stuart Raby, Saul Ramos-Sanchez, Michael Ratz,
  Patrick K.~S. Vaudrevange, and Akin Wingerter.
\newblock {A Mini-landscape of exact MSSM spectra in heterotic orbifolds}.
\newblock {\em Phys. Lett. B}, 645:88--94, 2007.

\bibitem{Lebedev:2007hv}
Oleg Lebedev, Hans~Peter Nilles, Stuart Raby, Saul Ramos-Sanchez, Michael Ratz,
  Patrick K.~S. Vaudrevange, and Akin Wingerter.
\newblock {The Heterotic Road to the MSSM with R parity}.
\newblock {\em Phys. Rev. D}, 77:046013, 2008.

\bibitem{Lebedev:2008un}
Oleg Lebedev, Hans~Peter Nilles, Saul Ramos-Sanchez, Michael Ratz, and Patrick
  K.~S. Vaudrevange.
\newblock {Heterotic mini-landscape. (II). Completing the search for MSSM vacua
  in a Z(6) orbifold}.
\newblock {\em Phys. Lett. B}, 668:331--335, 2008.

\bibitem{Blaszczyk:2009in}
Michael Blaszczyk, Stefan Groot~Nibbelink, Michael Ratz, Fabian Ruehle, Michele
  Trapletti, and Patrick K.~S. Vaudrevange.
\newblock {A Z2xZ2 standard model}.
\newblock {\em Phys. Lett. B}, 683:340--348, 2010.

\bibitem{Kappl:2010yu}
Rolf Kappl, Bjoern Petersen, Stuart Raby, Michael Ratz, Roland Schieren, and
  Patrick K.~S. Vaudrevange.
\newblock {String-Derived MSSM Vacua with Residual R Symmetries}.
\newblock {\em Nucl. Phys. B}, 847:325--349, 2011.

\bibitem{ParticleDataGroup:2016lqr}
C.~Patrignani et~al.
\newblock {Review of Particle Physics}.
\newblock {\em Chin. Phys. C}, 40(10):100001, 2016.

\bibitem{Kobayashi:2004ud}
Tatsuo Kobayashi, Stuart Raby, and Ren-Jie Zhang.
\newblock {Constructing 5-D orbifold grand unified theories from heterotic
  strings}.
\newblock {\em Phys. Lett. B}, 593:262--270, 2004.

\bibitem{Kobayashi:2004ya}
Tatsuo Kobayashi, Stuart Raby, and Ren-Jie Zhang.
\newblock {Searching for realistic 4d string models with a Pati-Salam symmetry:
  Orbifold grand unified theories from heterotic string compactification on a
  Z(6) orbifold}.
\newblock {\em Nucl. Phys. B}, 704:3--55, 2005.

\bibitem{Kobayashi:2006wq}
Tatsuo Kobayashi, Hans~Peter Nilles, Felix Ploger, Stuart Raby, and Michael
  Ratz.
\newblock {Stringy origin of non-Abelian discrete flavor symmetries}.
\newblock {\em Nucl. Phys. B}, 768:135--156, 2007.

\bibitem{Ko:2007dz}
Pyungwon Ko, Tatsuo Kobayashi, Jae-hyeon Park, and Stuart Raby.
\newblock {String-derived D(4) flavor symmetry and phenomenological
  implications}.
\newblock {\em Phys. Rev. D}, 76:035005, 2007.
\newblock [Erratum: Phys.Rev.D 76, 059901 (2007)].

\bibitem{T2K:2019bcf}
K.~Abe et~al.
\newblock {Constraint on the matter\textendash{}antimatter symmetry-violating
  phase in neutrino oscillations}.
\newblock {\em Nature}, 580(7803):339--344, 2020.
\newblock [Erratum: Nature 583, E16 (2020)].

\bibitem{deGiorgi:2022xhr}
Arturo de~Giorgi, Luca Merlo, and Stefan Pokorski.
\newblock {The Low-Scale Seesaw Solution to the $M_W$ and $(g-2)_\mu$
  Anomalies}.
\newblock 11 2022.

\bibitem{Belfatto:2023tbv}
Benedetta Belfatto and Sokratis Trifinopoulos.
\newblock {The remarkable role of the vector-like quark doublet in the Cabibbo
  angle and $W$-mass anomalies}.
\newblock 2 2023.

\bibitem{Baer:2016hfa}
Howard Baer, Vernon Barger, Hasan Serce, and Xerxes Tata.
\newblock {Natural generalized mirage mediation}.
\newblock {\em Phys. Rev. D}, 94(11):115017, 2016.

\bibitem{Hamaguchi:2022byw}
Koichi Hamaguchi, Natsumi Nagata, Genta Osaki, and Shih-Yen Tseng.
\newblock {Probing new physics in the vector-like lepton model by lepton
  electric dipole moments}.
\newblock {\em JHEP}, 01:100, 2023.

\bibitem{Jegerlehner:2009ry}
Fred Jegerlehner and Andreas Nyffeler.
\newblock {The Muon g-2}.
\newblock {\em Phys. Rept.}, 477:1--110, 2009.

\bibitem{Dermisek:2022hgh}
Radovan Dermisek.
\newblock {Muon g-2 and Other Observables in Models with Extended Higgs and
  Matter Sectors${}^{\#}$}.
\newblock {\em Moscow Univ. Phys. Bull.}, 77(2):102--107, 2022.

\bibitem{Athron:2021iuf}
Peter Athron, Csaba Bal\'azs, Douglas H.~J. Jacob, Wojciech Kotlarski, Dominik
  St\"ockinger, and Hyejung St\"ockinger-Kim.
\newblock {New physics explanations of a$_\mu$ in light of the FNAL muon g
  \ensuremath{-} 2 measurement}.
\newblock {\em JHEP}, 09:080, 2021.

\bibitem{Lavoura:2003xp}
L.~Lavoura.
\newblock {General formulae for f(1) ---\ensuremath{>} f(2) gamma}.
\newblock {\em Eur. Phys. J. C}, 29:191--195, 2003.

\bibitem{Hisano:1995cp}
J.~Hisano, T.~Moroi, K.~Tobe, and Masahiro Yamaguchi.
\newblock {Lepton flavor violation via right-handed neutrino Yukawa couplings
  in supersymmetric standard model}.
\newblock {\em Phys. Rev. D}, 53:2442--2459, 1996.

\bibitem{Ishiwata:2013gma}
Koji Ishiwata and Mark~B. Wise.
\newblock {Phenomenology of heavy vectorlike leptons}.
\newblock {\em Phys. Rev. D}, 88(5):055009, 2013.

\bibitem{Abada:2014kba}
A.~Abada, Manuel~E. Krauss, W.~Porod, F.~Staub, A.~Vicente, and Cedric Weiland.
\newblock {Lepton flavor violation in low-scale seesaw models: SUSY and
  non-SUSY contributions}.
\newblock {\em JHEP}, 11:048, 2014.

\bibitem{Dedes:2007ef}
Athanasios Dedes, Howard~E. Haber, and Janusz Rosiek.
\newblock {Seesaw mechanism in the sneutrino sector and its consequences}.
\newblock {\em JHEP}, 11:059, 2007.

\bibitem{Okada:1999zk}
Yasuhiro Okada, Ken-ichi Okumura, and Yasuhiro Shimizu.
\newblock {Mu --\ensuremath{>} e gamma and mu --\ensuremath{>} 3 e processes
  with polarized muons and supersymmetric grand unified theories}.
\newblock {\em Phys. Rev. D}, 61:094001, 2000.

\bibitem{Kuno:1999jp}
Yoshitaka Kuno and Yasuhiro Okada.
\newblock {Muon decay and physics beyond the standard model}.
\newblock {\em Rev. Mod. Phys.}, 73:151--202, 2001.

\bibitem{Altmannshofer:2014cfa}
Wolfgang Altmannshofer, Stefania Gori, Maxim Pospelov, and Itay Yavin.
\newblock {Quark flavor transitions in $L_\mu-L_\tau$ models}.
\newblock {\em Phys. Rev. D}, 89:095033, 2014.

\bibitem{Buras:2012jb}
Andrzej~J. Buras, Fulvia De~Fazio, and Jennifer Girrbach.
\newblock {The Anatomy of Z' and Z with Flavour Changing Neutral Currents in
  the Flavour Precision Era}.
\newblock {\em JHEP}, 02:116, 2013.

\bibitem{ParticleDataGroup:2018ovx}
M.~Tanabashi et~al.
\newblock {Review of Particle Physics}.
\newblock {\em Phys. Rev. D}, 98(3):030001, 2018.

\bibitem{Altmannshofer:2017wqy}
Wolfgang Altmannshofer, Christoph Niehoff, and David~M. Straub.
\newblock {$B_s\to\mu^+\mu^-$ as current and future probe of new physics}.
\newblock {\em JHEP}, 05:076, 2017.

\bibitem{Bobeth:2013uxa}
Christoph Bobeth, Martin Gorbahn, Thomas Hermann, Mikolaj Misiak, Emmanuel
  Stamou, and Matthias Steinhauser.
\newblock {$B_{s,d} \to l^+ l^-$ in the Standard Model with Reduced Theoretical
  Uncertainty}.
\newblock {\em Phys. Rev. Lett.}, 112:101801, 2014.

\bibitem{Buras:2002wq}
Andrzej~J. Buras, Piotr~H. Chankowski, Janusz Rosiek, and Lucja Slawianowska.
\newblock {Correlation between $\Delta M_s$ and $B^0_{s, d} \to \mu^{+}
  \mu^{-}$ in supersymmetry at large $\tan \beta$}.
\newblock {\em Phys. Lett. B}, 546:96--107, 2002.

\bibitem{Buras:2004ub}
Andrzej~J. Buras, Robert Fleischer, Stefan Recksiegel, and Felix Schwab.
\newblock {Anatomy of prominent B and K decays and signatures of CP violating
  new physics in the electroweak penguin sector}.
\newblock {\em Nucl. Phys. B}, 697:133--206, 2004.

\bibitem{Martin:1993zk}
Stephen~P. Martin and Michael~T. Vaughn.
\newblock {Two loop renormalization group equations for soft supersymmetry
  breaking couplings}.
\newblock {\em Phys. Rev. D}, 50:2282, 1994.
\newblock [Erratum: Phys.Rev.D 78, 039903 (2008)].

\bibitem{Babu:1993qv}
K.~S. Babu, Chung~Ngoc Leung, and James~T. Pantaleone.
\newblock {Renormalization of the neutrino mass operator}.
\newblock {\em Phys. Lett. B}, 319:191--198, 1993.

\bibitem{Antusch:2001ck}
Stefan Antusch, Manuel Drees, J\"orn Kersten, Manfred Lindner, and Michael
  Ratz.
\newblock {Neutrino mass operator renormalization revisited}.
\newblock {\em Phys. Lett. B}, 519:238--242, 2001.

\bibitem{Antusch:2001vn}
Stefan Antusch, Manuel Drees, J\"orn Kersten, Manfred Lindner, and Michael
  Ratz.
\newblock {Neutrino mass operator renormalization in two Higgs doublet models
  and the MSSM}.
\newblock {\em Phys. Lett. B}, 525:130--134, 2002.

\bibitem{Huang:2020hdv}
Guo-yuan Huang and Shun Zhou.
\newblock {Precise Values of Running Quark and Lepton Masses in the Standard
  Model}.
\newblock {\em Phys. Rev. D}, 103(1):016010, 2021.

\bibitem{68}
Adam Falkowski, David~M. Straub, and Avelino Vicente.
\newblock {Vector-like leptons: Higgs decays and collider phenomenology}.
\newblock {\em JHEP}, 05:092, 2014.

\bibitem{CMS:2022cik}
{Search for pair production of vector-like quarks in leptonic final states at
  $\sqrt{s}=13~\mathrm{TeV}$}.
\newblock 2022.

\bibitem{Alonso:2017uky}
Rodrigo Alonso, Peter Cox, Chengcheng Han, and Tsutomu~T. Yanagida.
\newblock {Flavoured $B-L$ local symmetry and anomalous rare $B$ decays}.
\newblock {\em Phys. Lett. B}, 774:643--648, 2017.

\bibitem{Allanach:2019mfl}
B.~C. Allanach, J.~M. Butterworth, and Tyler Corbett.
\newblock {Collider constraints on Z$^\prime$ models for neutral current
  B-anomalies}.
\newblock {\em JHEP}, 08:106, 2019.

\bibitem{Bonilla:2017lsq}
Cesar Bonilla, Tanmoy Modak, Rahul Srivastava, and Jose W.~F. Valle.
\newblock {$U(1)_{B_3-3L_\mu}$ gauge symmetry as a simple description of $b\to
  s$ anomalies}.
\newblock {\em Phys. Rev. D}, 98(9):095002, 2018.

\bibitem{Kohda:2018xbc}
Masaya Kohda, Tanmoy Modak, and Abner Soffer.
\newblock {Identifying a $Z'$ behind $b \to s \ell \ell$ anomalies at the LHC}.
\newblock {\em Phys. Rev. D}, 97(11):115019, 2018.

\bibitem{Wells:1997ag}
James~D. Wells.
\newblock {Mass density of neutralino dark matter}.
\newblock {\em Adv. Ser. Direct. High Energy Phys.}, 21:269--287, 2010.

\bibitem{Hisano:2017spq}
Junji Hisano, Takumi Kuwahara, Yuji Omura, and Takeki Sato.
\newblock {Two-loop Anomalous Dimensions for Four-Fermi Operators in
  Supersymmetric Theories}.
\newblock {\em Nucl. Phys. B}, 922:77--93, 2017.

\bibitem{Baer:2012cf}
Howard Baer, Vernon Barger, Peisi Huang, Dan Mickelson, Azar Mustafayev, and
  Xerxes Tata.
\newblock {Radiative natural supersymmetry: Reconciling electroweak fine-tuning
  and the Higgs boson mass}.
\newblock {\em Phys. Rev. D}, 87(11):115028, 2013.

\bibitem{Endo:2011mc}
Motoi Endo, Koichi Hamaguchi, Sho Iwamoto, and Norimi Yokozaki.
\newblock {Higgs Mass and Muon Anomalous Magnetic Moment in Supersymmetric
  Models with Vector-Like Matters}.
\newblock {\em Phys. Rev. D}, 84:075017, 2011.

\end{thebibliography}

% Please avoid comments such as "For a review'', "For some examples",
% "and references therein" or move them in the text. In general,
% please leave only references in the bibliography and move all
% accessory text in footnotes.

% Also, please have only one work for each \bibitem.

\end{document}